\documentclass[%
 aip,
 amsmath,amssymb,
 reprint,%
]{revtex4-1}

\usepackage[utf8]{inputenc}
\usepackage[T1]{fontenc}
\usepackage{mathptmx}
\usepackage{etoolbox}

\usepackage{graphicx}% Include figure files
\usepackage{dcolumn}% Align table columns on decimal point
\usepackage{bm}% bold math
\usepackage{physics}
\usepackage{multirow}

\usepackage{color,soul}
\usepackage{xspace}
\usepackage{xcolor}
\usepackage{siunitx}
\usepackage[version=4]{mhchem}

\makeatletter
\def\@email#1#2{%
 \endgroup
 \patchcmd{\titleblock@produce}
  {\frontmatter@RRAPformat}
  {\frontmatter@RRAPformat{\produce@RRAP{*#1\href{mailto:#2}{#2}}}\frontmatter@RRAPformat}
  {}{}
}%
\makeatother
\begin{document}

\preprint{AIP/123-QED}

\title{Generic Maximum-Valence Model for Fluid Polyamorphism}

\author{Nikolay A. Shumovskyi}%
\affiliation{Department of Physics, Boston University, Boston, MA 02215, USA}

\author{Sergey V. Buldyrev}%
\email{buldyrev@yu.edu}

\affiliation{Department of Physics, Boston University, Boston, MA 02215, USA}
\affiliation{ Department of Physics, Yeshiva University, New York, NY 10033, USA    \\}

\date{\today}% It is always \today, today,
             %  but any date may be explicitly specified

\begin{abstract}
Recently, maximal valence model has been proposed to model liquid-liquid phase transition induced by polymerization in sulfur. In this paper we present a simple generic model to describe liquid polyamorphism in single-component fluids using a maximum-valence approach for any arbitrary coordination number. The model contains three types of interactions:  i) atoms attract each other by van der Waals forces that generate a liquid-gas transition at low pressures, ii) atoms may form covalent bonds that induce association, and iii) additional repulsive forces between atoms with maximal valence and atoms with any valence. This additional repulsion generates liquid-liquid phase separation and the region of negative heat expansion coefficient (density anomaly) on a P-T phase diagram. We show the existence of liquid-liquid phase transitions for dimerization, polymerization, gelation and network formation for corresponding coordination numbers $z=1,2,..6$ and discuss the limits of this generic model for producing fluid polyamorphism.
\end{abstract}

\maketitle

The existence of two alternative liquid phases in a single-component substance is known as ``liquid polyamorphism'' \cite{Stanley_Liquid_2013,Anisimov2018,Tanaka_Liquid_2020}. A substance may be found to be polyamorphic by experimentally or computationally detecting a liquid-liquid phase transition (LLPT), which can be terminated at a liquid-liquid critical point (LLCP) \cite{Franzese2001,Sciorino_Silicon_2011}. Liquid polyamorphism has been observed in a variety of substances including: hydrogen \cite{Morales_H_2010,Zaghoo_H_2016,Simpson_H_2016}, helium \cite{Vollhardt_He_1990,Schmitt_He_2015}, sulfur \cite{Henry2020}, phosphorous \cite{Katayama2000,Katayama_Phos2_2004} and liquid carbon \cite{Glosli_Liquid_1999}, while being proposed to exist in selenium and tellurium \cite{Brazhkin_PT_1999,Plasienka_Structural_2015}. It has also been hypothesized in metastable deeply supercooled water below the temperature of spontaneous ice nucleation \cite{Stanley_Liquid_2013,Anisimov2018,Tanaka_Liquid_2020,Holten_Liquid_2012,Gallo2016,Duska_Water_2020,Caupin_Thermodynamics_2019,Poole1992,Holten2001,Debenedetti2020,Biddle_Two_2017,Debenedetti_One_1998}.

The phenomenon of liquid polyamorphism can be understood through the interconversion of the two alternative molecular or supramolecular states via a reversible reaction \cite{Anisimov2018,Longo2021,Caupin2021}. While for some polyamorphic systems, like supercooled water, this approach is still being debated, there are substances (such as hydrogen, sulfur, phosphorous, and liquid carbon) where liquid-liquid phase separation is indeed induced by a chemical reaction. For example, it was recently discovered that high-density sulfur, well above the liquid-gas critical pressure (in the range from $0.5$ to $\SI{2.0}{\giga\pascal}$), exhibits a LLPT indicated by a discontinuity in density from a low-density-liquid (LDL) monomer-rich phase to a high-density-liquid (HDL) polymer-rich phase \cite{Henry2020}. This liquid-liquid transition is found in a polymerized state of sulfur (observed above $\SI{160}{\degreeCelsius}$ at ambient pressure \cite{Sauer_Lambda_1967,Bellissent_Sulfur_1994,Kozhevnikov_Sulfur_2004,Tobolsky_Sulfur_1959,Tobolsky_Selenium_1960}). 
%However, with further increase of temperature, as the system approaches the liquid-gas phase transition (LGPT), the polymer chains gradually dissociate.
Another liquid-liquid transition accompanied by a reaction has been observed in hydrogen at extremely high-pressures (above \SI{325}{\giga\pascal} at ambient temperature \cite{Simpson_H_2016}), in which liquid-molecular hydrogen (dimers) dissociates into atomistic-metallic hydrogen \cite{Morales_H_2010,Zaghoo_H_2016}.

In this work, motivated by the recent discoveries of the LLPT in hydrogen \cite{Simpson_H_2016}, and continuing our previous work on the maximal valence model for sulfur \cite{Shumovskyi2022}, we propose a simple generic model to describe liquid polyamorphism in a variety of chemically-reacting fluids. The model combines the ideas of two-state thermodynamics \cite{Anisimov2018,Holten2001} with the maximum-valence approach \cite{Zaccarelli2005,Debenedetti,Debenedetti2}, in which atoms may form covalent bonds via a reversible reaction, changing their state according to their bond number. By mimicking the valence structure by maximum bond number, $z$, our model predicts the LLPT in systems with dimerization ($z=1$), polymerization ($z=2$), and gelation ($z>2$). We show that when the atoms with maximal valence repel atoms with any valence, phase separation is coupled to dimerization ($z=1$), polymerization ($z=2$), and gelation ($z>2$), thus generating the LLPT in polyamorphic substances. The key difference of this paper and the
previously published one\cite{Shumovskyi2022} is that here we investigate the case of \textit{repulsion} between atoms with the maximum valence z and any other atoms which causes the segregation of the atoms with the maximum valence  into the \textit{low} density phase; while in the previous paper in which we have attempted to model sulfur we investigate the case of \textit{attraction} of the atoms of maximal valence $(z=2)$ to each other which causes segregation of the polymerized atoms into the \textit{high} density phase.  Thus we have two classes of maximal valence models - one with attraction and another with repulsion, which drastically differ from each other. The models with repulsion studied here do have a density anomaly region with negative heat expansion coefficient $\alpha_P<0$, while the models with attraction do not.

\begin{figure*}[t]
\centering
{
\includegraphics[width=.7\textwidth]{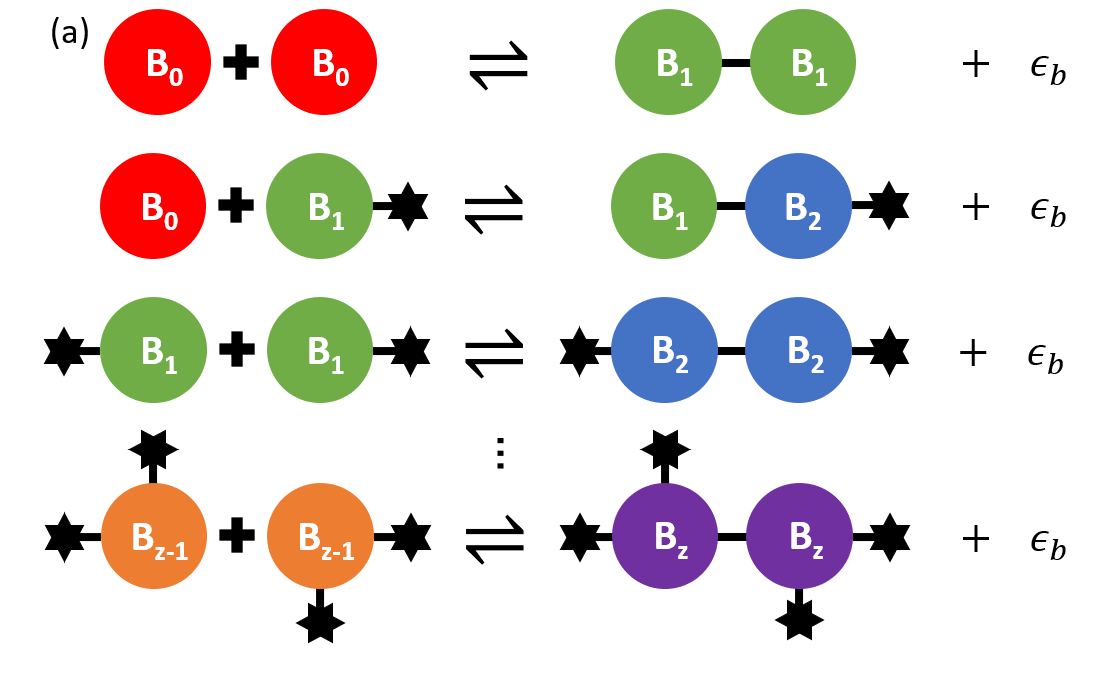}
}
\includegraphics[width=0.32\linewidth]{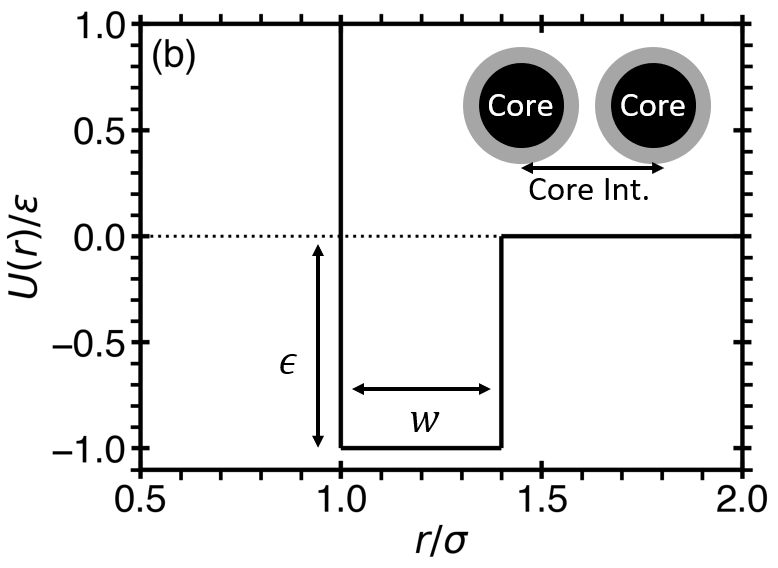}
\includegraphics[width=0.32\linewidth]{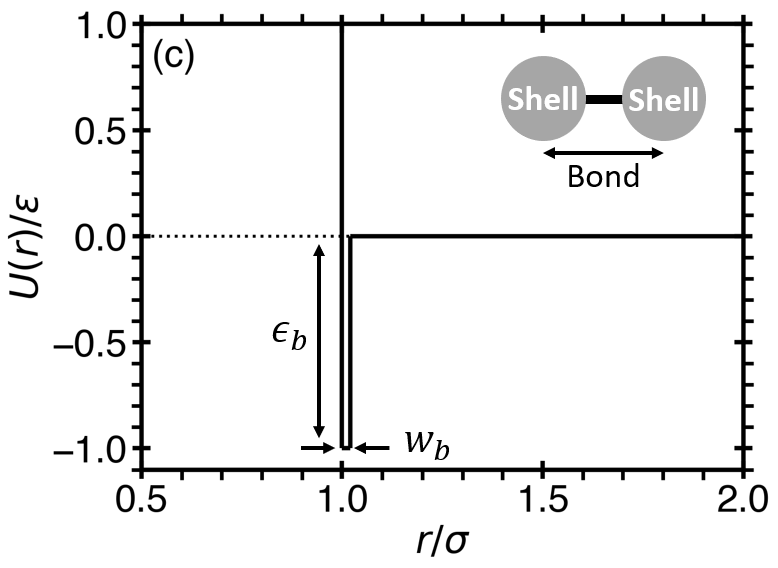}
\includegraphics[width=0.32\linewidth]{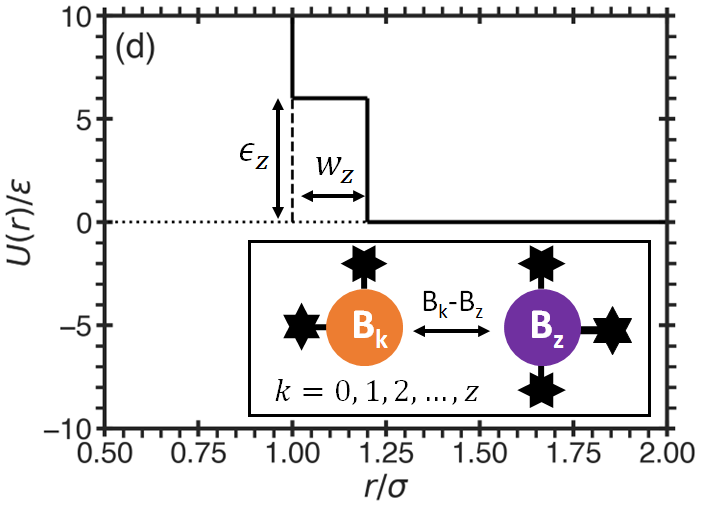}
\caption{Reactions and interactions in the generic maximum-valence model with repulsion. (a) $z(z+1)/2$ types of covalent bond-forming reversible chemical reactions that may occur in the system. If two atoms without bonds (B$_0$) collide with each other, they may form a bond and become B$_1$ atoms. If a B$_0$ and B$_1$ atom collide, they may form a bond and become B$_1$ and B$_2$ atoms, respectively. If two B$_1$ atoms collide with each other, they form an additional bond and become B$_2$ atoms. This continues until the atoms reach their maximum valency - state B$_z$. (b-d) The three major interactions between atoms, in which each atom is composed of a core and shell, both with a diameter $\sigma$ and mass $m$. $U(r)$ is the pair potential energy and $r$ is the distance from the centers of the particles. (b) The cores of each atom interact with an attractive square well of depth $\epsilon$ and width $w$. (c) The shells may react to form covalent bonds that consist of a narrow well with depth $\epsilon_{b} $ and width $w_{b}$. (d) Phase segregation is coupled to dimerization, polymerization, gelation, etc., via the additional repulsive interactions between atoms in state B$_z$ and atoms in a state B$_k$ with $k\leq z$, described by a square shoulder of height $-\epsilon_z$ and width $w_{z}$.
}
\label{fig1}
\end{figure*}

\section{Maximum-Valence Model} We model LLPT induced by molecular interconversion in polyamorphic substances by characterizing each atom by its coordination number $k\leq z$, the number of bonds it has with other atoms. Depending on the coordination number, each atom is assigned to distinguished $z+1$ states: B$_0$ (with zero bonds), B$_1$ (with one bond), B$_k$ (with $k$ bonds), and finally B$_z$ (with $z$ bonds). Atoms cannot form more than $z$ bonds and, consequently, will associated into either dimers, for $z=1$, or linear polymers, $z=2$, or some network structure $z>2$. All of the atoms in the system may change their state by forming or breaking a covalent bond via a reversible reaction. Fig.~\ref{fig1}a depicts all $z(z+1)/2$ types of reversible reactions that may occur in the system. In this work, we demonstrate that the minimum ingredients required to produce a LLPT are the following: i) the van der Waals interactions between atoms, which produce a LGPT; ii) covalent bonds between atoms, which induce association; and iii), as we hypothesize, additional repulsive interactions between atoms with maximum valence ($k = z$) and atoms with any valence ($k \leq z$), that are needed to couple phase segregation to dimerization, polymerization or gelation. These three ingredients are illustrated by square-well potentials in Figs.~\ref{fig1}(b-d). 

%In case of sulfur, the additional attraction between atoms in neighboring chains may stem from the fact that in real polymers the covalent bond is shorter than the diameter of the unbonded (``free'') atoms, such that the attractive wells of bonded atoms in neighboring chains overlap with each other \cite{Stell1972,Stillinger1993,Jagla2001,Franzese2001,Gibson2006,Skibinsky2004}. This effectively creates an additional zone of attraction between polymer chains, which is a common attribute that produces LLPTs in soft-core potentials \cite{Jagla2001,Franzese2001}.

In case of hydrogen, the pioneering quantum calculations of Wigner and Huntington\cite{Wigner1935} suggested that at high density the energy of the metallic lattice is lower than the energy of molecular lattice. In other words, at high densities H$_2$-dimers disassociate due to a steric effect, i.e. electron shells of H$_2$-dimers are getting larger than intermolecular distances. This steric repulsion is expected to happen not only for the case of hydrogen, but for other molecular liquids with higher valence.

Similar rule may be applied to the case of water. It was found experimentally that at high densities when the fifth neighbors are being pushed into the first coordination shell of a molecule, hydrogen bonds bifurcate and the energy of the bifurcated bonds are roughly half of the energy of the straight bonds \cite{Giguere1984}. Essentially this implies that the shells of the atoms with coordination number four become impenetrable for a fifth intruder at low temperatures.

%In case of hydrogen, the additional repulsive interactions between bonded atoms and non-bonded atoms mimic the fact that the atoms in the molecules H$_2$ have a larger repulsive diameter than atoms in metallic state which lacks bonded electrons. 

%Without this potential, with characteristic energy $\epsilon_{zz}$, and consequently, in the absence of associated atoms, no LLPT will occur. We note that this simplification is in the spirit of common semi-phenomenological models of non-ideal binary mixtures, such as the Flory-Huggins theory of polymer solutions \cite{Flory_Polymer_1941,Huggins_Solutions_1941,Rey_PIPS_1996,Luo_PIPS_2006} or a regular-solution model \cite{Hildebrand_Regular_1962}.

To verify our hypothesis, we implement these three ingredients of interactions via an event-driven MD technique \cite{Alder1959,Rapaport2004}; in particular, we use a discrete MD package (DMD) that only includes particles interacting through spherically-symmetric step-wise potentials, which may form bonds via reversible reactions \cite{Buldyrev_Application_2008}. We simulate an NVT ensemble of $N=1000$ atoms in a cubic box with periodic boundaries at various constant densities and temperatures. The temperature is controlled by a Berendsen thermostat \cite{Berendsen1984}. The van der Waals and covalent-bonding interactions are implemented by separating each atom into two overlapping hard spheres (a core and a shell), with the same diameter $\sigma$ and mass $m$, see Figs.~\ref{fig1}(b-d). The connection between the core and its shell is represented by an infinite square-well potential of width $d\ll\sigma$. The cores and shells of different atoms do not interact with each other. The core represents the atom without its valence electrons. It interacts with other cores via a wide potential well with depth $\epsilon$ and width $w \sim \sigma$. We use $\sigma$, $m$, and $\epsilon$
as units of length, mass, and energy, respectively, and measure all other  physical quantities using combination of these units. For example, temperature $T$ is measured in units of $\epsilon/k_B$; pressure, $P$ is measured in units of $\epsilon/\sigma^3$; and time $t$ is measured in units of $\sigma\sqrt{m/\epsilon}$ . We note that all physical parameters reported below are normalized by the appropriate combination of mass $m$, length
$\sigma$, and energy $\epsilon$ units, as used in Ref. \cite{Skibinsky2004} %We note, that for large values $z>2$, these interactions are unnecessary for the existence of the LGPT, which exists between a gaseous state of low valence atoms with $k\leq 1$ and a condensed state $k\geq 2$.  
Meanwhile, the shell represents the outer valence electron cloud. It can form bonds with other shells via a narrow potential well with depth $\epsilon_b = \epsilon$ and width $w_b$ (Fig.~\ref{fig1}c), which models the breaking and forming of covalent bonds. 
When the shell interactions are included and the system may form covalent bonds, the location of the LGCP changes, but not significantly. Note that for $z\geq 3$, the liquid-gas critical point may form  due to bonds only without wdV attraction between the cores as in the original maximal valence model\cite{Zaccarelli2005}. Also we introduce an additional repulsive potential shoulder (with depth $\epsilon_{z}$ and width $w_{z}$, Fig.~\ref{fig1}d) to model the steric repulsion of the atoms with $z$ bonds, which are not chemically bonded to each other. Here we assume that $\epsilon_z<0$ means repulsion, while the $\epsilon_z>0$ means attraction. We also emphasize the key difference between the case of repulsion in this work and attraction for the atoms of maximal valence model for sulfur\cite{Shumovskyi2022}. In case of attraction the shells in the state $B_z$ attract each other, but do not interact with other shells other than by hard-core repulsion with diameter $\sigma$. Conversely, for the case of repulsion shells in the state $B_z$ repel from each other and from all other shells,
while other shells do not interact with each other except by forming and breaking bonds, with the condition that two non-bonded atoms with $k<z$ cannot be within distance from each other smaller than the bond length $w_b$. Technically, this is achieved by assigning to such a pair a very strong repulsive potential at distance $w_b$. 
\smallbreak

%We note that during either the formation or breaking of a bond, the new state of the reacting particles may modify the potential energy of the interactions with their non-bonded neighboring particles \cite{Buldyrev_Application_2008}.

We note that during either the formation or breaking of
a bond, the new state of the reacting particles may modify
the potential energy of their interactions with their non-bonded
neighboring particles. In our model, this occurs when particles in the state B$_k$ convert to the state B$_{k-1}$ (or vice versa). To maintain the conservation of energy, we calculate the change of the total potential energy, $\Delta U$, due to the change of the state of the reacting particles and subtract it from the kinetic energy of the reacting pair. As a consequence, the equations for computing the new velocities \cite{Buldyrev_Application_2008} may not have real solutions. In this case, the bond will not form or break, and the reacting particles will conserve their states through an elastic collision.
\smallbreak
%In this work, we introduce the generic maximum-valence model to calculate the phase behavior in a variety of chemically-reacting systems exhibiting liquid polyamorphism by mimicking their valence structures and bond formation through the maximum bond number, $z$. 
The applicability of the maximum-valence model has been already tested for the case of sulfur ($z=2$, $\epsilon_z >0$)\cite{Shumovskyi2022}. Here we will focus on an arbitrary value of $z$ and additional repulsive interactions $\epsilon_z<0$. Note that for the case of attraction ($\epsilon_z >0$) the high density phase is polymerized. In the case of repulsion $\epsilon_z<0$, the low density phase is dimerized, polymerized or forms network with coordination number $z$. By tuning $z$, the phase behavior of a variety of substances can be described. For instance, for $z=1$, dimerization-induced phase separation (such as in high-pressure hydrogen \cite{Morales_H_2010,Zaghoo_H_2016,Simpson_H_2016}) is investigated, while for $z>2$, gelation-induced phase separation is investigated. In particular, the phase behavior of more complex chemically-reacting systems, such as phosphorous ($z=3$) \cite{Katayama2000,Katayama_Phos2_2004} and supercooled water, forming hydrogen instead of covalent bonds, ($z>3$)$^{18-32,34,35}$, can be investigated.

\section{Dimerization ($z=1$)}
We start with the simplest case of $z=1$, i.e. dimerization. We are allowing our system to have only two types of atoms, B$_0$ and B$_1$, which interact and may react by the scheme described by Fig. 1 with the maximal valence $z=1$. We found that in the case of repulsion between the atoms with the maximal valence, B$_1$, we see liquid-liquid phase transition produced by dimerization of atoms while reducing pressure.
To mimic the behavior of the hydrogen
we select the following set of parameters $w=0.5$, $w_b=0.1$, $\epsilon_b=6$, $w_{1}=0.2$, $\epsilon_{1}=-12$. The choice of a very large values of $\epsilon_b$ and $\epsilon_z$, comparatively to $\epsilon$ of the van der Waals forces is motivated by the desire to make the temperature and pressure of the liquid gas critical point to be much smaller than the temperature and pressure of the liquid-liquid critical point. For the selected set of parameters, the LGCP is located at $T_\text{c}^\text{LG}=0.91\pm 0.02$, $P_\text{c}^\text{LG}=0.030\pm 0.002$, $\rho_\text{c}^\text{LG}=0.225\pm 0.005$. Fig. 2a shows the $P-T$ phase diagram for this set of parameters near the LLCP with the crossing of isochores at $T_\text{c}^\text{LL}=2.12$, $P_\text{c}^\text{LL}=8.1655$, $\rho_\text{c}^\text{LL}=0.5775$, the liquid-liquid coexistence line shown with blue line ending in the LLCP marked with a red ellipse. Note, that in the region near the critical point the slopes of the isochores become negative which means that there exists a region of density anomaly in which the density of liquid decreases upon cooling. Moreover, it may indicate that the critical point is located in the region of density anomaly which implies that the critical isochores and the liquid-liquid coexistence line have a negative slope. Fig. 2b shows the apparent van der Waals loops on the $P-\rho$ phase diagram, which correspond to the Liquid-liquid phase transition.

\begin{figure}[t]
    \centering

    \includegraphics[width=\linewidth]{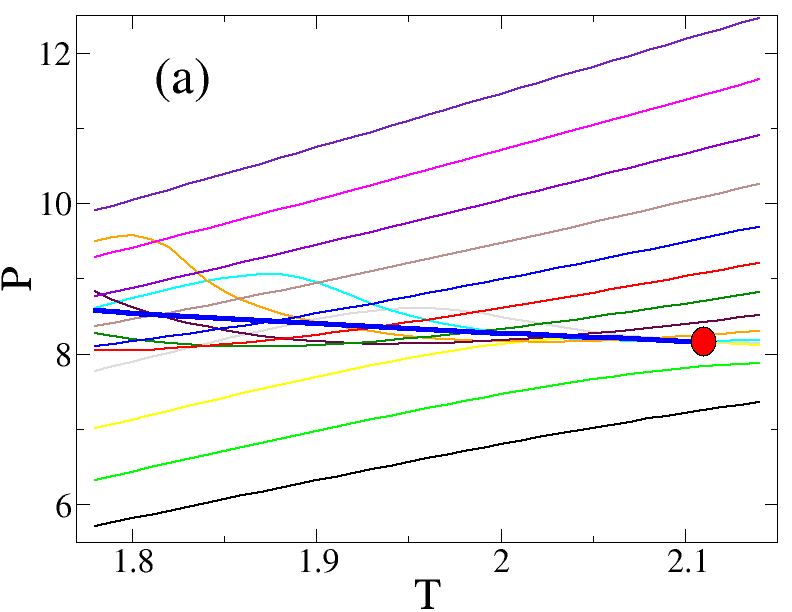}
    \includegraphics[width=\linewidth]{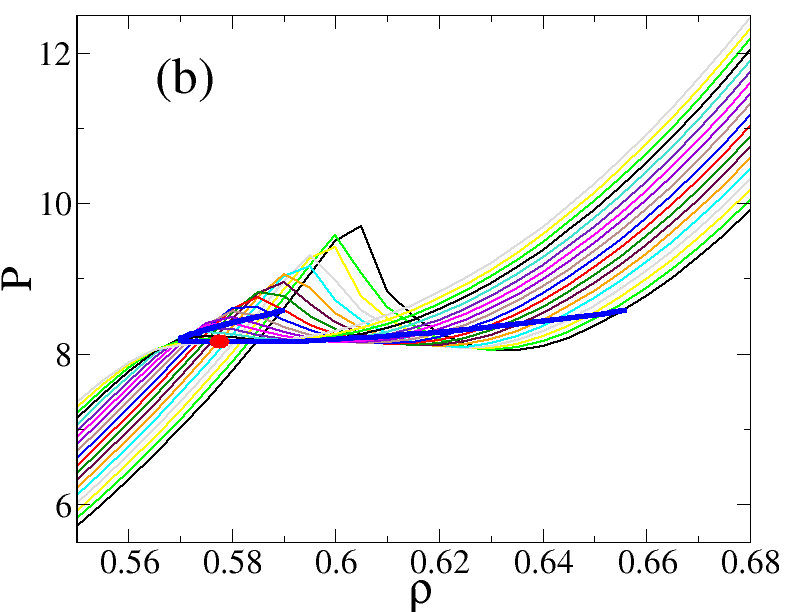}
    \caption{Phase diagrams for the maximum-valence model of dimerization, $z=1$, ($w_{b}=0.10$) obtained in an NVT ensemble after $t=10^6$ time units. (a) The isochores in the $P-T$ plane withe $\rho=0.550-0.680$ in steps $\Delta\rho =0.005$.
    (b) The isotherms in the $P$-$\rho$ plane with $T=1.78-2.14$ in steps $\Delta T=0.01$. In both figures, the liquid-liquid coexistence curves are calculated via the Maxwell construction and indicated by the blue curves. The liquid-liquid ($T_\text{c}^\text{LL}=2.12$, $P_\text{c}^\text{LL}=8.1655$, $\rho_\text{c}^\text{LL}=0.5775$) critical point is indicated by the red circles.}
    \label{Fig_Hydrogen_Phase_Dia}
\end{figure}

Another interesting feature of the model is found by looking at the temperature-density phase diagram, as well as the plot of temperature vs fraction of bonded atoms $\phi=N_1/N$, where $N$ is the total number of atoms and $N_1$ is the number of atoms in state $B_1$. In these diagrams we used reduced variables $\phi_r = (\phi-\phi_c)/\phi_c$, $T_r = (T-T_c)/T_c$, and $\rho_r = (\rho-\rho_c)/\rho_c$ with $\rho_c= 0.5775$, $T_c = 2.12$, $\phi_c = 0.350$ for $w_b = 0.10$ and $\rho_c= 0.76$, $T_c = 2.14434$, $\phi_c = 0.578$ for $w_b = 0.06$. Figure 3 shows that while the coexistence line on the $T_r$ - $\rho_r$ phase diagram is highly skewed to the right, the coexistence line on the $T_r$ - $\phi_r$ phase diagram is quite symmetric, which resembles the actual liquid-liquid coexistence line in hydrogen \cite{Fried2022}. Note, that one might apply two-state thermodynamics approach and compare the exact solution for hydrogen with our results from the simulations. This comparison is one of our future works in progress which requires a thorough analysis of the hydrogen system.

\begin{figure}[t]
    \centering
    \includegraphics[width=\linewidth]{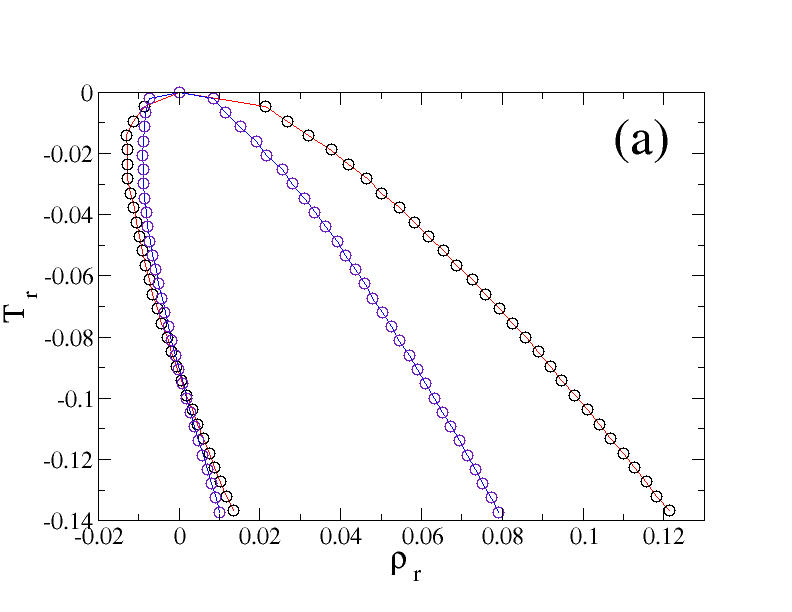}
    \includegraphics[width=\linewidth]{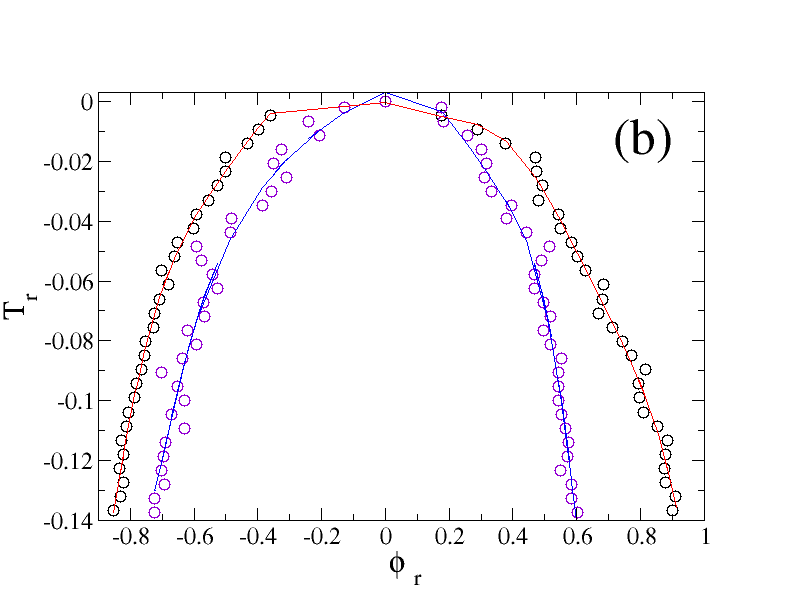}
    \caption{(a) $T_r$ - $\rho_r$ phase diagram for the maximum-valence model for dimerization, $z=1$, (with $w_{b}=0.10$, red,  and $w_b=0.06$, blue) obtained in an NVT ensemble after $t=10^6$ time units. (b) The reduced temperature dependence of the reduced fraction of atoms with one bond, $\phi_r$, in two coexisting liquid phases.}
    \label{Fig_Hydrogen_Phase_Dia}
\end{figure}

Next, we studied the dependence of the phase diagram on the parameters: $\epsilon_{1}$ and $w_{b}$, the strength of the additional interaction between the atoms in states with maximal valence, i.e. B$_1$ with z = 1, as well as the width of the bonds. As can be seen in the Fig. 4, upon increasing the strength $\epsilon_{1}$ all the LLCP parameters decrease - critical temperature, critical pressure, as well as critical density. Physically it means that upon increasing the repulsive energy between the bonded atoms, we effectively make the bonds less stable, and hence the temperature must be reduced to increase stability of the bonds necessary for the liquid-liquid phase segregation. As the repulsive energy between dimers increases the dimers get less penetrable and their density as well as the pressure, required to create such a density, drop. Thus we expect that both the density and the pressure of the second critical point decrease, as the repulsive energy increases. Fig 4(d) shows that we can force our system to have a negative slope of a liquid-liquid coexistence line on a $P-T$ phase diagram by increasing $|\epsilon_1|$, which qualitatively reproduces the negative slope in real hydrogen. Another important note is that upon increasing further the interaction energy between atoms with maximal valence, all four plots reach the plateau which corresponds to the limit $|\epsilon_1|/\epsilon_b \to  \infty$, when the repulsion strength between B$_1$ atoms is much larger than bond strength and the system reaches the limit when the dimers become effectively impenetrable, so the system behaves like a system of quasi-hard spheres made of dimers which have larger effective radius rather than hydrogen monomers, and the LGCP parameters and the slope of the coexistence line rapidly acquire the values corresponding to this limiting case, because the probability to find an atom sitting on a repulsive shoulder decreases according to Arhenius law $\exp[\epsilon_1/(k_B T)]$ . Indeed, all four parameters approach their limit exponentially with $\epsilon_1\to -\infty$. The limit $\epsilon_1 \to -\infty$ corresponds to the impenetrable electron shells of the $H_2$ molecules, which cannot exist if another atom enters their electron shells and must break if such an event happens.  

\begin{figure}[t]
    \centering
    \includegraphics[width=0.49\linewidth]{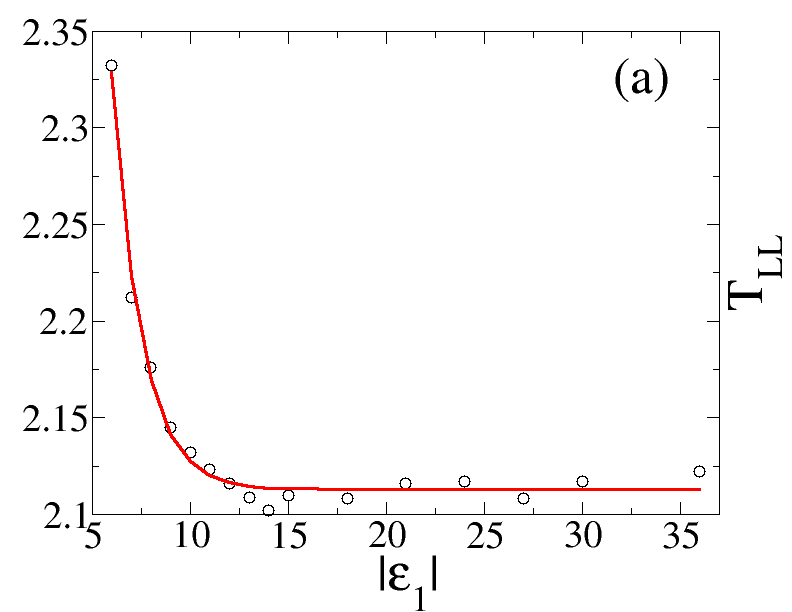}
    \includegraphics[width=0.49\linewidth]{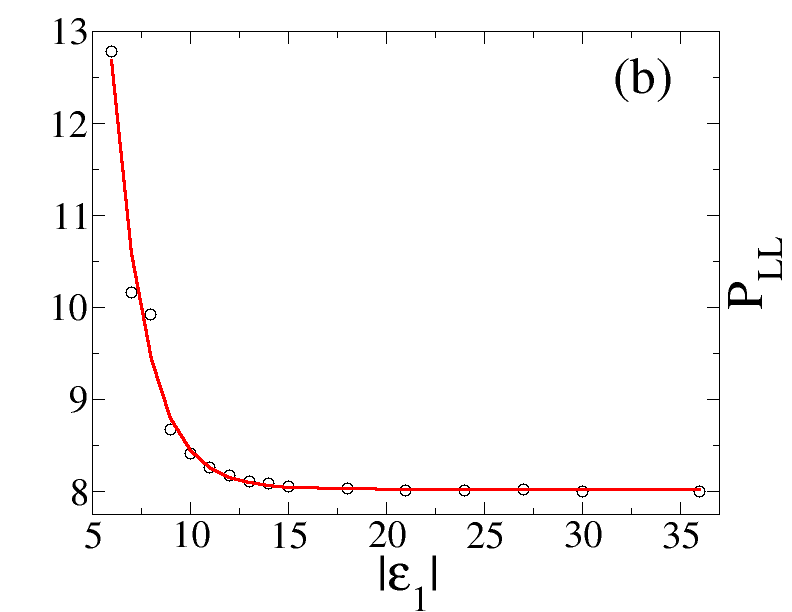}
    \includegraphics[width=0.49\linewidth]{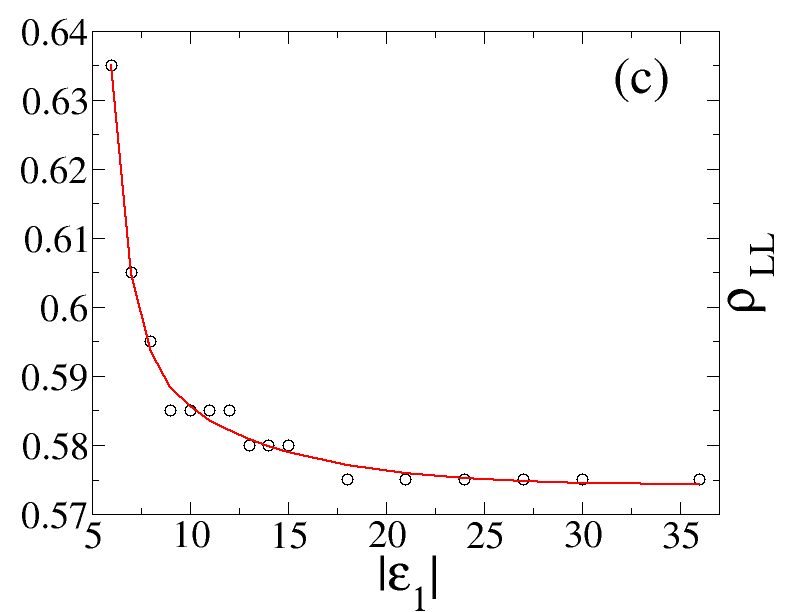}
    \includegraphics[width=0.49\linewidth]{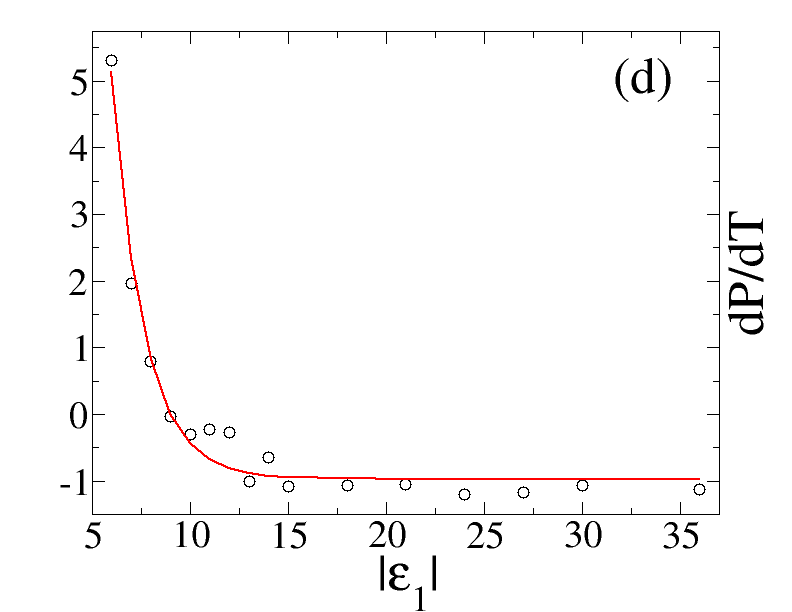}
    \caption{The dependence of the Liquid-liquid critical point parameters on the additional interaction strength $|\epsilon_1|$ for $z=1$, $w=0.5$, $d=0.1$, $w_b=0.1$, $w_1=0.2$, $\epsilon_b=6$}
    
\end{figure}

Figure 5 shows the dependence of liquid-liquid critical pressure and critical density on the bond width at $\epsilon_b=6$, $\epsilon_1=-12$, keeping $w_1=2 w_b$:
as we increase the bond width, the system doesn't require that much pressure, or doesn't need to be in such a squeezed state, for monomers to start dimerization, so the LLCP moves down in both $P$ and $\rho$ axis, however, the temperature of LLCP stays approximately the same, since the increase of the bond width only plays the entropic effect, not the energetic one. We note that upon decreasing the bond width further below $w = 0.06$, the LLCP becomes submerged below the crystallization line, and the phase transition is not observed anymore. However, increasing the bond width further will only decrease the $(P_c,\rho_c)$ further, up to the point when the liquid-liquid critical pressure becomes negative or LLCP is getting destroyed by the uLG spinodal. The phase diagrams corresponding to $w_b = 0.06$ are show in Fig. 6.

Note, that when we compare the $T-\rho$ phase diagrams, as well as the plots of $T-\phi$, for the two values of the bond width $w_b = 0.10$ and $w_b=0.06$, we find that the the coexistence regions are indeed squeezed, when we reduce $w_b$ from $w_b=0.10$ to $w_b=0.06$. The percentage difference of the densities of the HDL and LDL  far away from the critical point is about 11.5\% for $w_b=0.10$ , while for $w_b=0.06$ it is 6.9\% (see Fig. 3). Note, that in hydrogen it is approximately 2\% \cite{Fried2022}. In theory, upon decreasing the bond width we should obtain the desired 2\%, however our simulations show that below $w_b=0.06$ the LLCP is submerged below the crystallization line.

The same effect can be achieved by reducing the width of repulsive interactions $w_1$, while keeping $w_b$ constant. The difference in densities decreases when $(w_1-w_b)/\sigma$ decreases, and reaches 6\% when $(w_1-w_b)/\sigma=0.06$, but further decrease leads to instantaneous crystallization in the region where the LLCP would be expected.

%By reducing $w_1$ while keeping $w_b=0.10$, we observe the same effect of reducing the percent difference between high- and low- density phases. However, when the ratio $(w_1-w_b)/w$ becomes less than 6\%, our system crystallizes and we cannot achieve density difference less than 6\%.

\begin{figure}[t]
\centering
{
\includegraphics[width=0.9\linewidth]{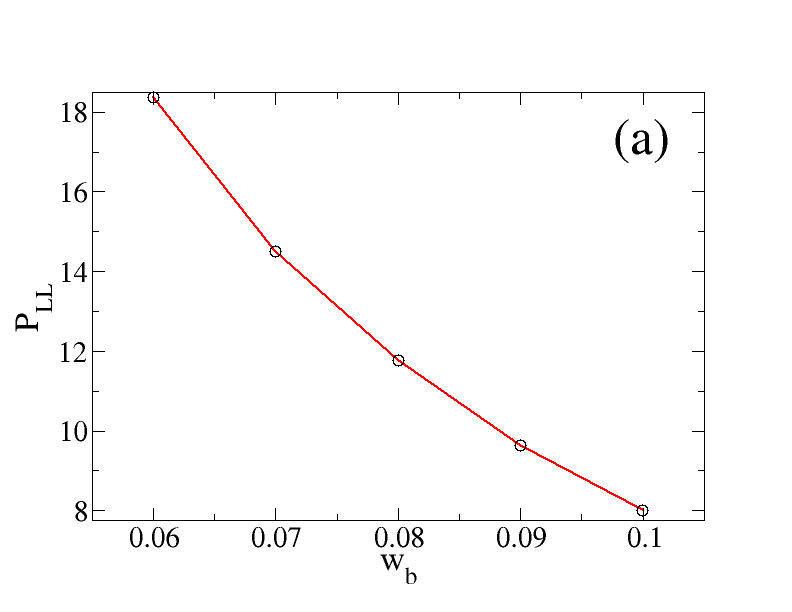}
\includegraphics[width=0.9\linewidth]{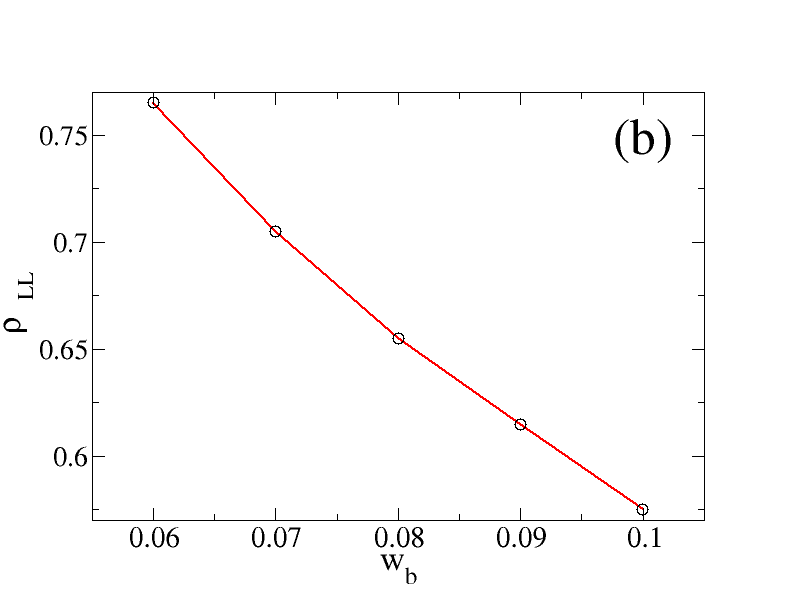}
}
\caption{The dependence of the liquid-liquid critical point parameters on the bond width for $z=1$, $\epsilon_1=-12$, $w=0.5$, $d=0.1$, $\epsilon_b=6$, $w_1=2w_b$ }
\label{fig2}
\end{figure}

\begin{figure}[t]
    \centering
    \includegraphics[width=\linewidth]{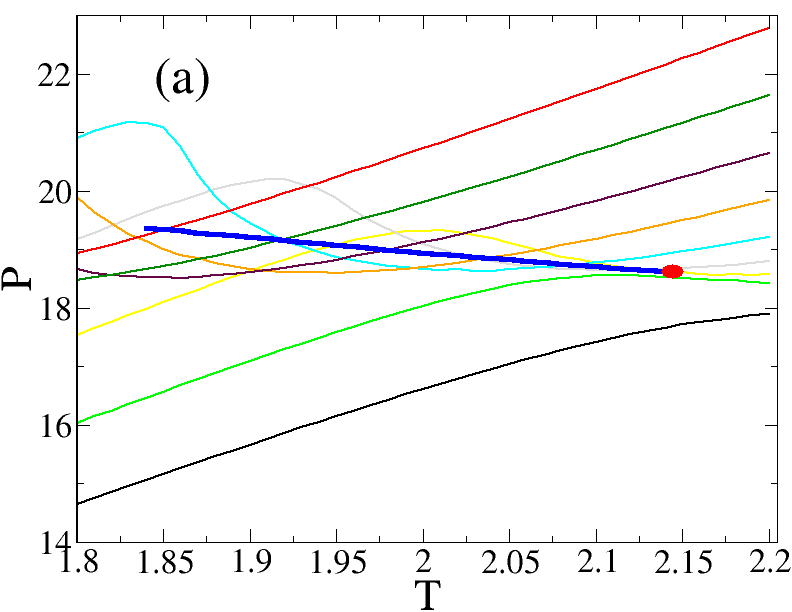}
    \includegraphics[width=\linewidth]{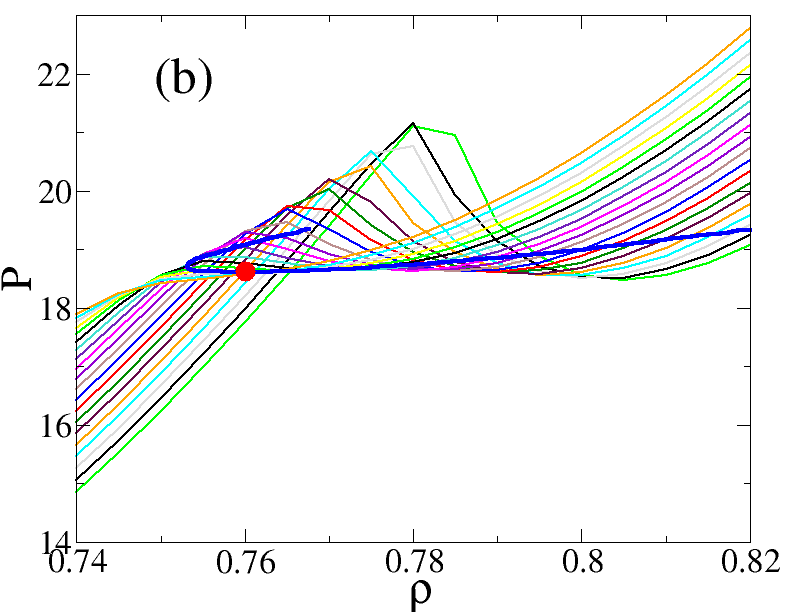}
    \caption{Phase diagrams for the maximum-valence model of dimerization, $z=1$, ($w_b=0.06$) obtained in an NVT ensemble after $t=10^6$ time units. (a) The isochores in the $P-T$ plane are $\rho=0.96-1.20$ for $\rho=0.760-0.820$ in steps $\Delta T=0.01$.
    (b) The isotherms in the $P$-$\rho$ plane fir $T=1.82-2.20$ in steps $\Delta T=0.01$. In both figures, the liquid-liquid coexistence curves are calculated via the Maxwell construction and indicated by the blue curves. The liquid-liquid  ($T_\text{c}^\text{LL}=2.14434$, $P_\text{c}^\text{LL}=18.618$, $\rho_\text{c}^\text{LL}=0.670$) critical point is indicated by the red circles. Other parameters are the same as in Fig.5}
    \label{Fig_Hydrogen_Phase_Dia}
\end{figure}

\section{Z = 2, 3, 4, 5, 6 (Polymerization, Gelation / Network Formation)} 
We can generalize our model for any arbitrary coordination number $z$. So far we have discussed in details only dimerization, $z=1$, and polymerization, $z=2$ for attractive interaction between B$_2$ atoms \cite{Shumovskyi2022}.
Upon studying other $z>2$, we found that our generic maximal valence model also produces liquid-liquid phase transitions induced by molecular interconversion in polyamorphic substances for higher $z$ as well, up to the point the critical density becomes so high that the LLCP moves below the crystallization line.

Figure 7 shows the dependence of critical parameters, critical temperature $T_c$, critical pressure $P_c$ and critical density $\rho_c$, on the coordination number $z$, while keeping all other parameters constant ($w=0.5$, $w_b=0.1$, $w_z=0.2$, $d=0.1$, $\epsilon_b=6$, $\epsilon_z=-6$). Upon increasing $z$ the liquid-liquid critical temperature decreases, while the liquid-gas critical temperature increases. Physically the temperature of LGCP increases upon increasing the coordination number $z$, because  the overall attraction increases. However, as $z$ increases both LG and LL critical pressures increase. We know that since $T^{LG}_c$ is increasing, it forces the pressure of liquid-gas also to increase. As for the density of LGCP, it only slightly increases for high $z$. On the other hand, the density of the LLCP increases from approximately 0.6 to 1. This is because the density of the low density phase which consists of atoms with coordination number $z$ increases with $z$ at a constant bond length. When $z=6$ the network density coincides with the density of the simple cubic lattice which has coordination number 6. For higher $z$ the LLPT becomes submerged below the crystallization line of the body-centered cubic lattice. The pressure needed to create such a density also increases, so we see that the LLCP pressure increases with $z$. The most intriguing observation of Fig.7 (a) is the apparent crossing of the temperatures of the  LGCP and LLCP as z increases. This is in agreement with the experimental observations that for hydrogen (z=1) the liquid gas critical temperature (33K) is much smaller than the hypothetical critical temperature of the liquid-liquid critical point (>400K), while for water (z=4) the liquid-gas critical temperature is 647K is mach larger than
the hypothetical critical point of the supercooled water $T<230K$.  
Note that the pressures for LGCP and LLCP are quite different, the later is orders of magnitude larger than the former, so the physical processes behind these critical points are quite independent. The crossing of the critical temperature curves at a particular $z$ is just a coincidence, an interplay of the parameters. What important is that $T_c^{LG}$ is increasing, while $T_c^{LL}$ is decreasing. While the increase of $T_c^{LG}$ is obvious because the low density liquid has a much lower energy because additional bonds for larger $z$ decrease the energy of the liquid phase, the decrease of $T_c^{LL}$ with $z$ can be explained by entropic effect. Increasing the number of neighbors in the coordination shell from $z-1$ to $z$ leads to a larger entropy loss for larger $z$.  This observation is consistent with the behavior of the slope of the LL coexistence versus $z$ (Fig. 8). Indeed, for all other parameters equal, the LL  coexistence line for $z=1$ has a positive slope on a PT diagram, while for $z>=2$ the slope is negative and is getting more negative for higher $z$. This means that the low density phase for $z=1$ (dimeric) has higher entropy than low density phase (monomeric). In contrast for $z=2$ the low density phase (polymeric) has lower entropy than the high density phase (dimeric) and for higher values of z this difference is getting more negative. If heat expansion coefficient is positive, density and entropy are negatively correlated, but for negative heat expansion (density anomaly) the density and entropy are positively correlated\cite{Gallo2016}. This implies that the higher density phase has more disordered neighborhood than the low density phase. And this disordering upon compression increases with $z$. Thus analyzing the slope of the coexistence line leads to the same conclusion, as analyzing the decrease of the critical temperature with $z$.

The choice of parameters for our model is quite simple and natural. In reality, a wide region in the parameter space produces a phase diagram with a region of density anomaly and a LLCP with a negatively sloped coexistence line, like it happens in water\cite{Poole1992}. In the previous section we find that for $z=1$, the slope of the coexistence line is negative only for large additional repulsive interaction for the dimerized atoms $\epsilon_1 \approx -2\epsilon_b$. For larger $z$ the the slope becomes more negative even for weaker repulsion $\epsilon_z=-\epsilon_b$ (Fig. 8). Our preliminary results show that increasing the width of the repulsive shoulder ($w_z$) to 1.35 at $z=4$  while keeping all other parameters constant as in Fig 7. reduces the pressure of the region of the density anomaly to zero like in water. In this case the critical point is located at negative pressures very close to the LG spinodal. Further increase of $w_z$ leads to the disappearance of the critical point below the LG spinodal, similar to the critical point free scenario\cite{Gallo2016,Poole1994}.

Reducing the width to 0.15 leads to the disappearance of the density anomaly together with the critical point. Increasing the height of the shoulder to infinity does not change the phase diagram after a certain point, approximately $\epsilon_z / \epsilon_b >2$, while the decrease of the height to $\epsilon_b/2$ eliminates the critical point. The detailed investigation of the phase diagram for the wide shoulder is very important for understanding the phase diagram of water and is the subject of our current work.

Lastly, Fig. 9 shows the radial distribution function, $g(r)$, and the structure factor, $S(q)$, for the system with a repulsive interaction for $z=4$ with a set of parameters $w=0.5$, $w_b=0.1$, $w_z=0.2$, $d=0.1$, $\epsilon_b=6$, $\epsilon_z=-6$ along the isochore $\rho = 0.87$, which crosses the line of the liquid-liquid phase transition illustrated by three snapshots at $T=1.35$ (high density phase), $T=1.09$ (coexistence), and $T=0.86$ (low density phase). On the $g(r)$ and $S(q)$ plots for low temperature one can see the splitting of the first shell into two peaks, the first peak corresponds to the atoms with maximal valence $(z =  4)$, while the second peak corresponds to the atoms which cannot penetrate into the repulsive shoulder. At low temperatures, we can see the low-density phase consisting predominantly of atoms with four bounds (white), while at high temperature we see the high density phase consisting of atoms with three bonds (blue) and with a small fraction of atoms with four bonds (white) and two bonds (green).
To relate this picture to the structure of water, we must emphasize that only straight hydrogen bonds are counted as bonds in the maximal valence model. The bifurcated bonds are counted as the intruders onto the repulsive shoulder with the correspondent energy penalty.

\begin{figure}[t]
\centering
{
\includegraphics[width=0.9\linewidth]{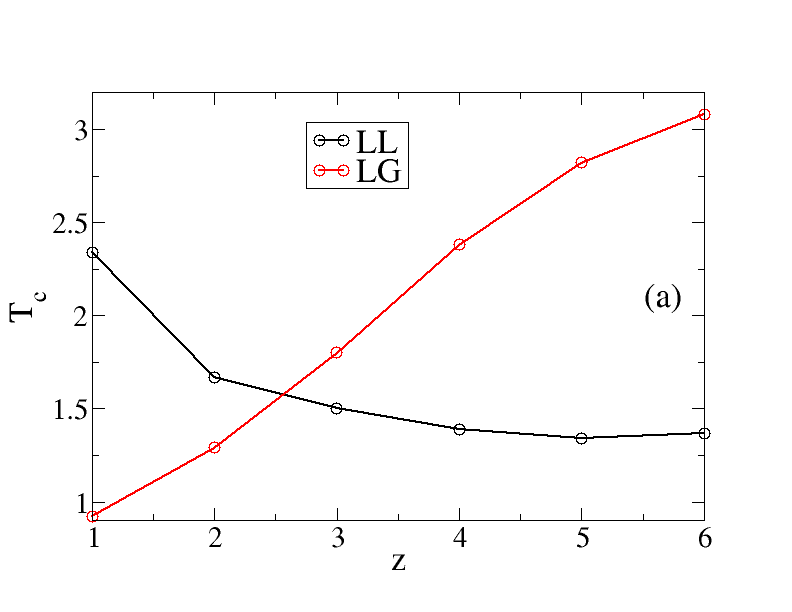}
\includegraphics[width=0.9\linewidth]{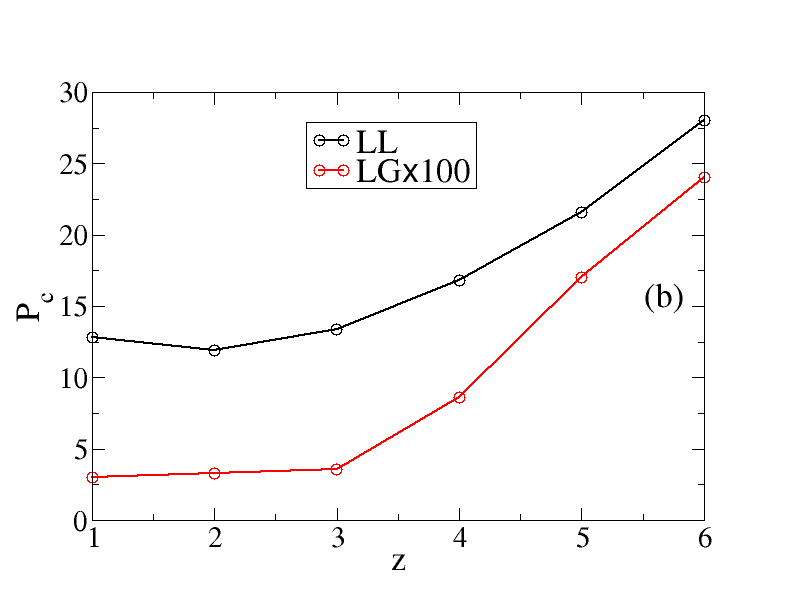}
\includegraphics[width=0.9\linewidth]{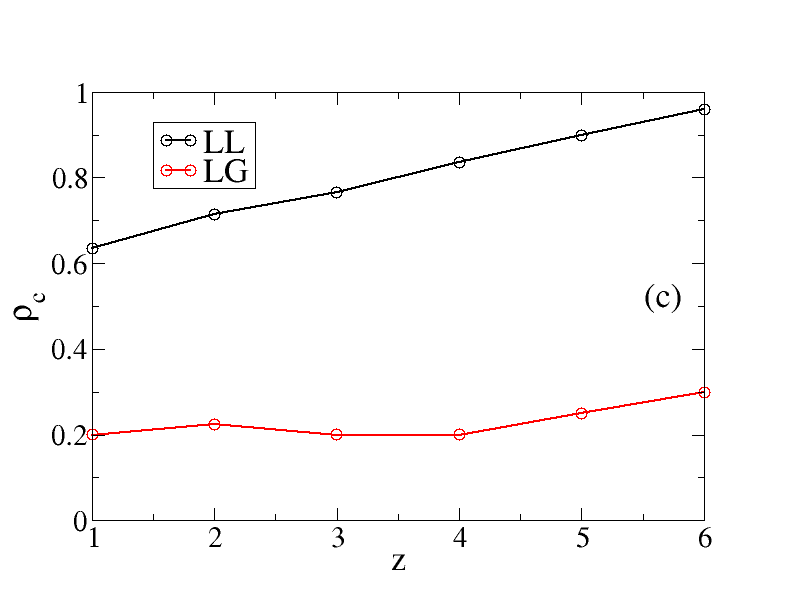}
}
\caption{The dependence of the Liquid-liquid critical point parameters on coordination number $z$. The parameters are: $w=0.5$, $w_b=0.1$, $w_z=0.2$, $d=0.1$, $\epsilon_b=6$, $\epsilon_z=-6$}
\label{fig6}
\end{figure}

%\subsection{General remarks on Z-dependence}

\begin{figure}[t]
\centering
{
\includegraphics[width=0.9\linewidth]{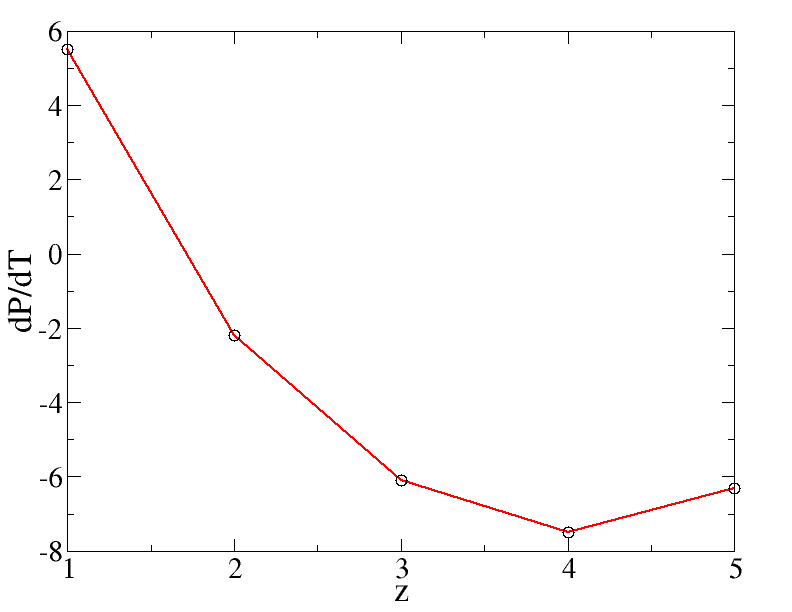}
}
\caption{The dependence of the slope of the liquid-liquid coexistence line on coordination number $z$. The parameters are: $w = 0.5$, $w_b = 0.1$,
$w_z = 0.2$, $d = 0.1$, $\epsilon_b = 6$, $\epsilon_z=-6$ }
\label{figH}
\end{figure}

\begin{figure}[t]
\centering
{
\includegraphics[width=0.9\linewidth]{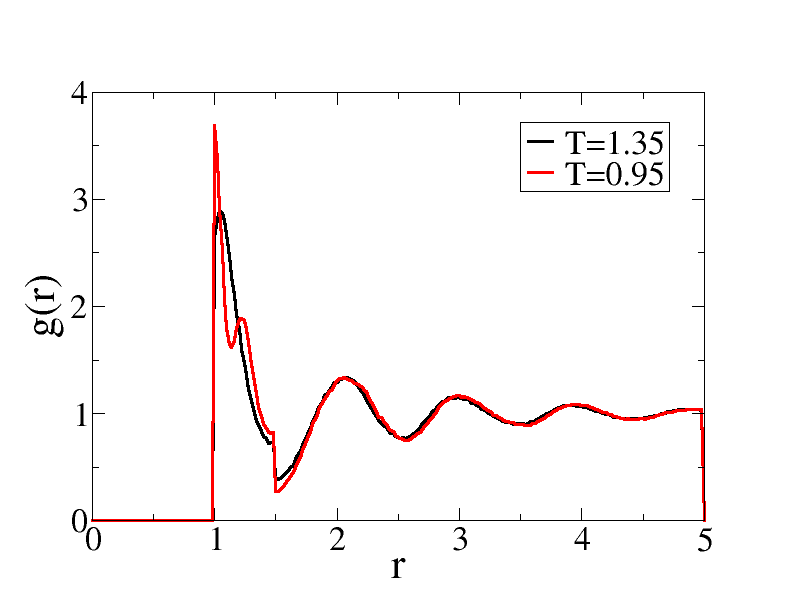}
\includegraphics[width=0.9\linewidth]{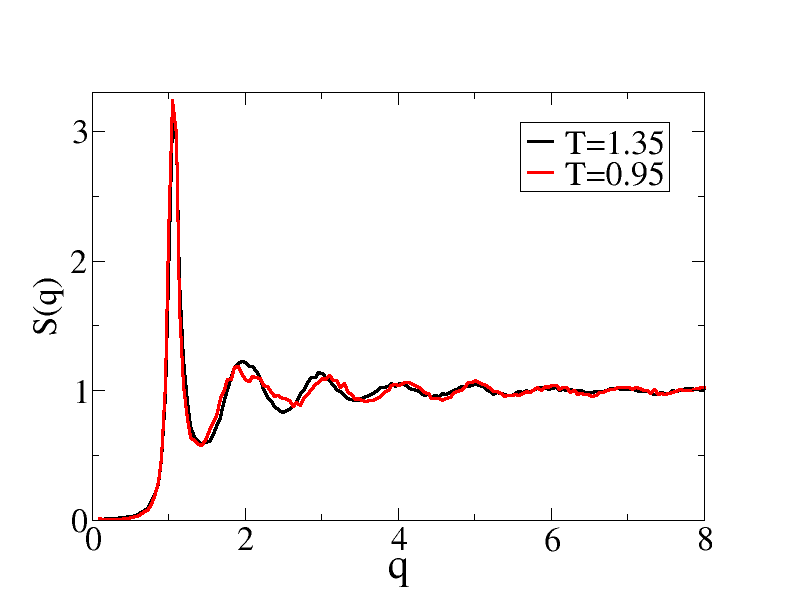}
\includegraphics[width=0.9\linewidth]{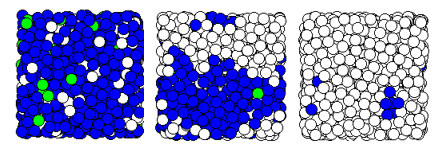}
}
\caption{Radial g-factor, $g(r)$, and structure factor, $S(q)$, for $z=4$ system with repulsion. The parameters are: $w=0.5$, $w_b=0.1$, $w_z=0.2$, $d=0.1$, $\epsilon_b=6$, $\epsilon_z=-6$. The snapshots of the system are presented for $T=1.35, T=1.09, T=0.86$ (left to right).}
\label{fig6}
\end{figure}

%\red{ The highest possible z is 12, when the spherical atoms must form the hexagonal close packed or the face centered cubic lattice. But already, for $z=6$, the structure of the liquid must resemble a simple cubic lattice, and the liquid becomes prone to crystallization, exactly as we observe. However, the liquid phase with $z=6$ can still be observed in a very narrow range of densities. For $z>6$ the liquid at high density, necessary to achieve this coordination number, spontaneously crystallizes into the body centered cubic lattice, which has coordination number eight.  We have included this comment into the revised manuscript.}

\section{Conclusion} 
The maximum-valence model describes liquid polyamorphism in a variety of chemically-reacting fluids. By tuning the maximum valency, $z$ (maximum coordination number), of the model, the liquid-liquid phase transitions (LLPT) in these systems can be investigated. We show that when the atoms with maximal valence $k=z$ repel atoms with valence $k\leq z$, the LLPT is generated by the coupling between phase separation and the chemical reactions. 

We showed that when $z=1$, the LLPT is induced by dimerization (e.g. hydrogen at extremely large pressures\cite{Simpson_H_2016}). We described how by tuning the size of the molecule relatively to the atomic hydrogen radius, our model may be used to reproduce the real hydrogen. By decreasing the difference between the bond length, $w_b$, and the repulsive range of the atoms in the molecule, $w_1$, one may reduce the difference between percentage of high- and low- density phases to 6\%, while in the real hydrogen this difference is reported to be approximately 2\% \cite{Fried2022}. After reducing the difference between the $w_b$ and $w_1$ even further, the system crystallizes. Since the exact position of the LLCP in hydrogen is still under debate\cite{Fried2022},
our results suggest that 2\% difference between high- and low- density phases is inconsistent with the existence of the LLPT above the crystallization line.

In our previous work for $z=2$, we showed that LLPT induced by polymerization could describe LLPT in sulfur\cite{Shumovskyi2022}. For $z\ge 3$, the LLPT could be induced either by gelation or by molecular network formation \cite{Zaccarelli2005}. For example, it could be used to model the phase behavior of liquid phosphorous with $z=3$ \cite{Katayama2000,Katayama_Phos2_2004} or supercooled water with $z>3$ \cite{Gallo2016,Debenedetti2}. Indeed for high values of $z$ we obtained the negative slope of the coexistence line on $P-T$ plane as well as the relationships between the critical parameters of the model, $T^{LG}_c \gg T^{LL}_c$ and $P^{LG}_c \ll P^{LL}_c$, which qualitatively reproduces the real water experiments \cite{Kim2020} and simulations of more accurate all-atom models. 

An interesting question is the existence of the liquid-liquid transition for very high values of z.
The highest possible z for short bonds is 12, when the spherical atoms must form the hexagonal close packed (hcp) or the face centered cubic (fcc) lattice. But already, for $z=6$, the structure of the liquid must resemble a simple cubic lattice, and the liquid becomes prone to crystallization.  However, the liquid phase with $z=6$ can still be observed in a very narrow range of densities. For $z>6$ the liquid at high densities, necessary to achieve these high coordination numbers, spontaneously crystallizes into the body centered cubic lattice (bcc), which has coordination number eight. Our preliminary observations suggests that these bcc crystals may display an isostructural solid-solid phase transition ending in solid-solid critical point\cite{Bolhius1984}.

In a future study, the two-state thermodynamics of liquid polyamorphism \cite{Anisimov2018, Caupin2021,Longo2021,Fried2022} could be applied to these systems to develop the equation of state, which would determine the anomalies of the physical properties in these systems, especially near the critical points.

\begin{acknowledgments}
The authors thank Miklail Anisimov, Fr\'ed\'eric Caupin, Pablo G. Debenedetti, Thomas Longo, Francesco Sciortino, and Eugene I. Shakhnovich for useful discussions. This work is supported by the National Science Foundation. The research at Boston University was supported by NSF Award No. 1856496. S.V.B. acknowledges the partial support of this research through Bernard W. Gamson Computational Science Center at Yeshiva College.
\end{acknowledgments}

%\section*{Data Availability Statement}
%The data that support the findings of this study are openly available in [repository name] at http://doi.org/[doi], reference number [reference number].

\section*{References}
\bibliography{refs}% Produces the bibliography via BibTeX.

%merlin.mbs aipnum4-1.bst 2010-07-25 4.21a (PWD, AO, DPC) hacked
%Control: key (0)
%Control: author (8) initials jnrlst
%Control: editor formatted (1) identically to author
%Control: production of article title (0) allowed
%Control: page (1) range
%Control: year (1) truncated
%Control: production of eprint (0) enabled
\begin{thebibliography}{47}%
\makeatletter
\providecommand \@ifxundefined [1]{%
 \@ifx{#1\undefined}
}%
\providecommand \@ifnum [1]{%
 \ifnum #1\expandafter \@firstoftwo
 \else \expandafter \@secondoftwo
 \fi
}%
\providecommand \@ifx [1]{%
 \ifx #1\expandafter \@firstoftwo
 \else \expandafter \@secondoftwo
 \fi
}%
\providecommand \natexlab [1]{#1}%
\providecommand \enquote  [1]{``#1''}%
\providecommand \bibnamefont  [1]{#1}%
\providecommand \bibfnamefont [1]{#1}%
\providecommand \citenamefont [1]{#1}%
\providecommand \href@noop [0]{\@secondoftwo}%
\providecommand \href [0]{\begingroup \@sanitize@url \@href}%
\providecommand \@href[1]{\@@startlink{#1}\@@href}%
\providecommand \@@href[1]{\endgroup#1\@@endlink}%
\providecommand \@sanitize@url [0]{\catcode `\\12\catcode `\$12\catcode
  `\&12\catcode `\#12\catcode `\^12\catcode `\_12\catcode `\%12\relax}%
\providecommand \@@startlink[1]{}%
\providecommand \@@endlink[0]{}%
\providecommand \url  [0]{\begingroup\@sanitize@url \@url }%
\providecommand \@url [1]{\endgroup\@href {#1}{\urlprefix }}%
\providecommand \urlprefix  [0]{URL }%
\providecommand \Eprint [0]{\href }%
\providecommand \doibase [0]{http://dx.doi.org/}%
\providecommand \selectlanguage [0]{\@gobble}%
\providecommand \bibinfo  [0]{\@secondoftwo}%
\providecommand \bibfield  [0]{\@secondoftwo}%
\providecommand \translation [1]{[#1]}%
\providecommand \BibitemOpen [0]{}%
\providecommand \bibitemStop [0]{}%
\providecommand \bibitemNoStop [0]{.\EOS\space}%
\providecommand \EOS [0]{\spacefactor3000\relax}%
\providecommand \BibitemShut  [1]{\csname bibitem#1\endcsname}%
\let\auto@bib@innerbib\@empty
%</preamble>
\bibitem [{\citenamefont {Stanely}(2013)}]{Stanley_Liquid_2013}%
  \BibitemOpen
  \bibfield  {author} {\bibinfo {author} {\bibfnamefont {H.~E.}\ \bibnamefont
  {Stanely}},\ }\href@noop {} {\emph {\bibinfo {title} {Liquid
  Polymorphism}}},\ edited by\ \bibinfo {editor} {\bibfnamefont {A.~R.~D.}\
  \bibnamefont {Stuart A.~Rice}},\ \bibinfo {series} {Advances in Chemical
  Physics}, Vol.\ \bibinfo {volume} {152}\ (\bibinfo  {publisher} {JohnWiley \&
  Sons},\ \bibinfo {year} {2013})\BibitemShut {NoStop}%
\bibitem [{\citenamefont {Anisimov}\ \emph {et~al.}(2018)\citenamefont
  {Anisimov}, \citenamefont {Duška}, \citenamefont {Caupin}, \citenamefont
  {Amrhein}, \citenamefont {Rosenbaum},\ and\ \citenamefont
  {Sadus}}]{Anisimov2018}%
  \BibitemOpen
  \bibfield  {author} {\bibinfo {author} {\bibfnamefont {M.~A.}\ \bibnamefont
  {Anisimov}}, \bibinfo {author} {\bibfnamefont {M.}~\bibnamefont {Duška}},
  \bibinfo {author} {\bibfnamefont {F.}~\bibnamefont {Caupin}}, \bibinfo
  {author} {\bibfnamefont {L.~E.}\ \bibnamefont {Amrhein}}, \bibinfo {author}
  {\bibfnamefont {A.}~\bibnamefont {Rosenbaum}}, \ and\ \bibinfo {author}
  {\bibfnamefont {R.~J.}\ \bibnamefont {Sadus}},\ }\bibfield  {title} {\enquote
  {\bibinfo {title} {Thermodynamics of fluid polyamorphism},}\ }\href {\doibase
  10.1103/PhysRevX.8.011004} {\bibfield  {journal} {\bibinfo  {journal} {Phys.
  Rev. X}\ }\textbf {\bibinfo {volume} {8}},\ \bibinfo {pages} {011004}
  (\bibinfo {year} {2018})}\BibitemShut {NoStop}%
\bibitem [{\citenamefont {Tanaka}(2020)}]{Tanaka_Liquid_2020}%
  \BibitemOpen
  \bibfield  {author} {\bibinfo {author} {\bibfnamefont {H.}~\bibnamefont
  {Tanaka}},\ }\bibfield  {title} {\enquote {\bibinfo {title} {Liquid–liquid
  transition and polyamorphism},}\ }\href {\doibase
  https://doi.org/10.1063/5.0021045} {\bibfield  {journal} {\bibinfo  {journal}
  {J. Chem. Phys.}\ }\textbf {\bibinfo {volume} {153}},\ \bibinfo {pages}
  {130901} (\bibinfo {year} {2020})}\BibitemShut {NoStop}%
\bibitem [{\citenamefont {Franzese}\ \emph {et~al.}(2001)\citenamefont
  {Franzese}, \citenamefont {Malescio}, \citenamefont {Skibinsky},
  \citenamefont {Buldyrev},\ and\ \citenamefont {Stanley}}]{Franzese2001}%
  \BibitemOpen
  \bibfield  {author} {\bibinfo {author} {\bibfnamefont {G.}~\bibnamefont
  {Franzese}}, \bibinfo {author} {\bibfnamefont {G.}~\bibnamefont {Malescio}},
  \bibinfo {author} {\bibfnamefont {A.}~\bibnamefont {Skibinsky}}, \bibinfo
  {author} {\bibfnamefont {S.~V.}\ \bibnamefont {Buldyrev}}, \ and\ \bibinfo
  {author} {\bibfnamefont {H.~E.}\ \bibnamefont {Stanley}},\ }\bibfield
  {title} {\enquote {\bibinfo {title} {Generic mechanism for generating a
  liquid-liquid phase transition},}\ }\href {\doibase
  https://doi.org/10.1038/35055514} {\bibfield  {journal} {\bibinfo  {journal}
  {Nature}\ }\textbf {\bibinfo {volume} {409}} (\bibinfo {year} {2001}),\
  https://doi.org/10.1038/35055514}\BibitemShut {NoStop}%
\bibitem [{\citenamefont {Sciortino}(2011)}]{Sciorino_Silicon_2011}%
  \BibitemOpen
  \bibfield  {author} {\bibinfo {author} {\bibfnamefont {F.}~\bibnamefont
  {Sciortino}},\ }\bibfield  {title} {\enquote {\bibinfo {title}
  {Liquid–liquid transitions: Silicon in silico},}\ }\href {\doibase
  https://doi.org/10.1038/nphys2038} {\bibfield  {journal} {\bibinfo  {journal}
  {Nat. Phys.}\ }\textbf {\bibinfo {volume} {7}},\ \bibinfo {pages} {523--524}
  (\bibinfo {year} {2011})}\BibitemShut {NoStop}%
\bibitem [{\citenamefont {Morales}\ \emph {et~al.}(2010)\citenamefont
  {Morales}, \citenamefont {Pierleoni}, \citenamefont {Schwegler},\ and\
  \citenamefont {Ceperley}}]{Morales_H_2010}%
  \BibitemOpen
  \bibfield  {author} {\bibinfo {author} {\bibfnamefont {M.~A.}\ \bibnamefont
  {Morales}}, \bibinfo {author} {\bibfnamefont {C.}~\bibnamefont {Pierleoni}},
  \bibinfo {author} {\bibfnamefont {E.}~\bibnamefont {Schwegler}}, \ and\
  \bibinfo {author} {\bibfnamefont {D.~M.}\ \bibnamefont {Ceperley}},\
  }\bibfield  {title} {\enquote {\bibinfo {title} {Evidence for a first-order
  liquid-liquid transition in high-pressure hydrogen from ab initio
  simulations},}\ }\href {\doibase 10.1073/pnas.1007309107} {\bibfield
  {journal} {\bibinfo  {journal} {Proc. Natl. Acad. Sci.}\ }\textbf {\bibinfo
  {volume} {107}},\ \bibinfo {pages} {12799--12803} (\bibinfo {year}
  {2010})}\BibitemShut {NoStop}%
\bibitem [{\citenamefont {Zaghoo}, \citenamefont {Salamat},\ and\ \citenamefont
  {Silvera}(2016)}]{Zaghoo_H_2016}%
  \BibitemOpen
  \bibfield  {author} {\bibinfo {author} {\bibfnamefont {M.}~\bibnamefont
  {Zaghoo}}, \bibinfo {author} {\bibfnamefont {A.}~\bibnamefont {Salamat}}, \
  and\ \bibinfo {author} {\bibfnamefont {I.~F.}\ \bibnamefont {Silvera}},\
  }\bibfield  {title} {\enquote {\bibinfo {title} {Evidence of a first-order
  phase transition to metallic hydrogen},}\ }\href {\doibase
  10.1103/PhysRevB.93.155128} {\bibfield  {journal} {\bibinfo  {journal} {Phys.
  Rev. B}\ }\textbf {\bibinfo {volume} {93}},\ \bibinfo {pages} {155128}
  (\bibinfo {year} {2016})}\BibitemShut {NoStop}%
\bibitem [{\citenamefont {Dalladay-Simpson}, \citenamefont {Howie},\ and\
  \citenamefont {Gregoryanz}(2016)}]{Simpson_H_2016}%
  \BibitemOpen
  \bibfield  {author} {\bibinfo {author} {\bibfnamefont {P.}~\bibnamefont
  {Dalladay-Simpson}}, \bibinfo {author} {\bibfnamefont {R.~T.}\ \bibnamefont
  {Howie}}, \ and\ \bibinfo {author} {\bibfnamefont {E.}~\bibnamefont
  {Gregoryanz}},\ }\bibfield  {title} {\enquote {\bibinfo {title} {Evidence for
  a new phase of dense hydrogen above 325 gigapascals},}\ }\href {\doibase
  https://doi.org/10.1038/nature16164} {\bibfield  {journal} {\bibinfo
  {journal} {Nature}\ }\textbf {\bibinfo {volume} {529}},\ \bibinfo {pages}
  {63--67} (\bibinfo {year} {2016})}\BibitemShut {NoStop}%
\bibitem [{\citenamefont {Vollhardt}\ and\ \citenamefont
  {Wölfle}(1990)}]{Vollhardt_He_1990}%
  \BibitemOpen
  \bibfield  {author} {\bibinfo {author} {\bibfnamefont {D.}~\bibnamefont
  {Vollhardt}}\ and\ \bibinfo {author} {\bibfnamefont {P.}~\bibnamefont
  {Wölfle}},\ }\href@noop {} {\emph {\bibinfo {title} {The Superfluid Phases
  of Helium 3}}}\ (\bibinfo  {publisher} {Taylor and Francis},\ \bibinfo
  {address} {London, UK},\ \bibinfo {year} {1990})\BibitemShut {NoStop}%
\bibitem [{\citenamefont {Schmitt}(2015)}]{Schmitt_He_2015}%
  \BibitemOpen
  \bibfield  {author} {\bibinfo {author} {\bibfnamefont {A.}~\bibnamefont
  {Schmitt}},\ }\bibfield  {title} {\enquote {\bibinfo {title} {Introduction to
  superfluidity},}\ \ }(\bibinfo  {publisher} {Springer International
  Publishing},\ \bibinfo {address} {Cham},\ \bibinfo {year} {2015})\BibitemShut
  {NoStop}%
\bibitem [{\citenamefont {Henry}\ \emph {et~al.}(2020)\citenamefont {Henry},
  \citenamefont {Mezouar}, \citenamefont {Garbarino}, \citenamefont {Sifré},
  \citenamefont {Weck},\ and\ \citenamefont {Datchi}}]{Henry2020}%
  \BibitemOpen
  \bibfield  {author} {\bibinfo {author} {\bibfnamefont {L.}~\bibnamefont
  {Henry}}, \bibinfo {author} {\bibfnamefont {M.}~\bibnamefont {Mezouar}},
  \bibinfo {author} {\bibfnamefont {G.}~\bibnamefont {Garbarino}}, \bibinfo
  {author} {\bibfnamefont {D.}~\bibnamefont {Sifré}}, \bibinfo {author}
  {\bibfnamefont {G.}~\bibnamefont {Weck}}, \ and\ \bibinfo {author}
  {\bibfnamefont {F.}~\bibnamefont {Datchi}},\ }\bibfield  {title} {\enquote
  {\bibinfo {title} {Liquid–liquid transition and critical point in
  sulfur},}\ }\href {\doibase https://doi.org/10.1038/s41586-020-2593-1}
  {\bibfield  {journal} {\bibinfo  {journal} {Nature}\ }\textbf {\bibinfo
  {volume} {584}},\ \bibinfo {pages} {382--386} (\bibinfo {year}
  {2020})}\BibitemShut {NoStop}%
\bibitem [{\citenamefont {Katayama}\ \emph {et~al.}(2000)\citenamefont
  {Katayama}, \citenamefont {Mizutani}, \citenamefont {Utsumi}, \citenamefont
  {Shimomura}, \citenamefont {Yamakata},\ and\ \citenamefont {ichi
  Funakoshi}}]{Katayama2000}%
  \BibitemOpen
  \bibfield  {author} {\bibinfo {author} {\bibfnamefont {Y.}~\bibnamefont
  {Katayama}}, \bibinfo {author} {\bibfnamefont {T.}~\bibnamefont {Mizutani}},
  \bibinfo {author} {\bibfnamefont {W.}~\bibnamefont {Utsumi}}, \bibinfo
  {author} {\bibfnamefont {O.}~\bibnamefont {Shimomura}}, \bibinfo {author}
  {\bibfnamefont {M.}~\bibnamefont {Yamakata}}, \ and\ \bibinfo {author}
  {\bibfnamefont {K.}~\bibnamefont {ichi Funakoshi}},\ }\bibfield  {title}
  {\enquote {\bibinfo {title} {A first-order liquid–liquid phase transition
  in phosphorus},}\ }\href {\doibase https://doi.org/10.1038/35003143}
  {\bibfield  {journal} {\bibinfo  {journal} {Nature}\ }\textbf {\bibinfo
  {volume} {403}},\ \bibinfo {pages} {170--173} (\bibinfo {year}
  {2000})}\BibitemShut {NoStop}%
\bibitem [{\citenamefont {Katayama}\ \emph {et~al.}(2004)\citenamefont
  {Katayama}, \citenamefont {Inamura}, \citenamefont {Mizutani}, \citenamefont
  {Yamakata}, \citenamefont {Utsumi},\ and\ \citenamefont
  {Shimomura}}]{Katayama_Phos2_2004}%
  \BibitemOpen
  \bibfield  {author} {\bibinfo {author} {\bibfnamefont {Y.}~\bibnamefont
  {Katayama}}, \bibinfo {author} {\bibfnamefont {Y.}~\bibnamefont {Inamura}},
  \bibinfo {author} {\bibfnamefont {T.}~\bibnamefont {Mizutani}}, \bibinfo
  {author} {\bibfnamefont {M.}~\bibnamefont {Yamakata}}, \bibinfo {author}
  {\bibfnamefont {W.}~\bibnamefont {Utsumi}}, \ and\ \bibinfo {author}
  {\bibfnamefont {O.}~\bibnamefont {Shimomura}},\ }\bibfield  {title} {\enquote
  {\bibinfo {title} {Macroscopic separation of dense fluid phase and liquid
  phase of phosphorus},}\ }\href {\doibase 10.1126/science.1102735} {\bibfield
  {journal} {\bibinfo  {journal} {Science}\ }\textbf {\bibinfo {volume}
  {306}},\ \bibinfo {pages} {848--851} (\bibinfo {year} {2004})}\BibitemShut
  {NoStop}%
\bibitem [{\citenamefont {Glosli}\ and\ \citenamefont
  {Ree}(1999)}]{Glosli_Liquid_1999}%
  \BibitemOpen
  \bibfield  {author} {\bibinfo {author} {\bibfnamefont {J.~N.}\ \bibnamefont
  {Glosli}}\ and\ \bibinfo {author} {\bibfnamefont {F.~H.}\ \bibnamefont
  {Ree}},\ }\bibfield  {title} {\enquote {\bibinfo {title} {Liquid-liquid phase
  transformation in carbon},}\ }\href {\doibase 10.1103/PhysRevLett.82.4659}
  {\bibfield  {journal} {\bibinfo  {journal} {Phys. Rev. Lett.}\ }\textbf
  {\bibinfo {volume} {82}},\ \bibinfo {pages} {4659--4662} (\bibinfo {year}
  {1999})}\BibitemShut {NoStop}%
\bibitem [{\citenamefont {Brazhkin}, \citenamefont {Popova},\ and\
  \citenamefont {Voloshin}(1999)}]{Brazhkin_PT_1999}%
  \BibitemOpen
  \bibfield  {author} {\bibinfo {author} {\bibfnamefont {V.~V.}\ \bibnamefont
  {Brazhkin}}, \bibinfo {author} {\bibfnamefont {S.~V.}\ \bibnamefont
  {Popova}}, \ and\ \bibinfo {author} {\bibfnamefont {R.~N.}\ \bibnamefont
  {Voloshin}},\ }\bibfield  {title} {\enquote {\bibinfo {title} {Pressure
  -temperature phase diagram of molten elements: selenium, sulfur and
  iodine},}\ }\href {\doibase https://doi.org/10.1016/S0921-4526(98)01318-0}
  {\bibfield  {journal} {\bibinfo  {journal} {Physica B}\ }\textbf {\bibinfo
  {volume} {265}},\ \bibinfo {pages} {64--71} (\bibinfo {year}
  {1999})}\BibitemShut {NoStop}%
\bibitem [{\citenamefont {Plašienka}, \citenamefont {Cifra},\ and\
  \citenamefont {Martoňák}(2015)}]{Plasienka_Structural_2015}%
  \BibitemOpen
  \bibfield  {author} {\bibinfo {author} {\bibfnamefont {D.}~\bibnamefont
  {Plašienka}}, \bibinfo {author} {\bibfnamefont {P.}~\bibnamefont {Cifra}}, \
  and\ \bibinfo {author} {\bibfnamefont {R.}~\bibnamefont {Martoňák}},\
  }\bibfield  {title} {\enquote {\bibinfo {title} {Structural transformation
  between long and short-chain form of liquid sulfur from ab initio molecular
  dynamics},}\ }\href {\doibase https://doi.org/10.1063/1.4917040} {\bibfield
  {journal} {\bibinfo  {journal} {J. Chem. Phys.}\ }\textbf {\bibinfo {volume}
  {142}},\ \bibinfo {pages} {154502--154512} (\bibinfo {year}
  {2015})}\BibitemShut {NoStop}%
\bibitem [{\citenamefont {Holten}\ and\ \citenamefont
  {Anisimov}(2012)}]{Holten_Liquid_2012}%
  \BibitemOpen
  \bibfield  {author} {\bibinfo {author} {\bibfnamefont {V.}~\bibnamefont
  {Holten}}\ and\ \bibinfo {author} {\bibfnamefont {M.~A.}\ \bibnamefont
  {Anisimov}},\ }\bibfield  {title} {\enquote {\bibinfo {title} {Entropy-driven
  liquid–liquid separation in supercooled water},}\ }\href {\doibase
  https://doi.org/10.1038/srep00713} {\bibfield  {journal} {\bibinfo  {journal}
  {Sci. Rep.}\ }\textbf {\bibinfo {volume} {2}},\ \bibinfo {pages} {713}
  (\bibinfo {year} {2012})}\BibitemShut {NoStop}%
\bibitem [{\citenamefont {Gallo}\ \emph {et~al.}(1994)\citenamefont {Gallo},
  \citenamefont {Amann-Winkel}, \citenamefont {Angell}, \citenamefont
  {Anisimov}, \citenamefont {Caupin}, \citenamefont {Chakravarty},
  \citenamefont {Lascaris}, \citenamefont {Loerting}, \citenamefont
  {Panagiotopoulos}, \citenamefont {Russo}, \citenamefont {Sellberg},
  \citenamefont {Stanley}, \citenamefont {Tanaka}, \citenamefont {Vega},
  \citenamefont {Xu},\ and\ \citenamefont {Pettersson}}]{Gallo2016}%
  \BibitemOpen
  \bibfield  {author} {\bibinfo {author} {\bibfnamefont {P.}~\bibnamefont
  {Gallo}}, \bibinfo {author} {\bibfnamefont {K.}~\bibnamefont {Amann-Winkel}},
  \bibinfo {author} {\bibfnamefont {C.~A.}\ \bibnamefont {Angell}}, \bibinfo
  {author} {\bibfnamefont {M.~A.}\ \bibnamefont {Anisimov}}, \bibinfo {author}
  {\bibfnamefont {F.}~\bibnamefont {Caupin}}, \bibinfo {author} {\bibfnamefont
  {C.}~\bibnamefont {Chakravarty}}, \bibinfo {author} {\bibfnamefont
  {E.}~\bibnamefont {Lascaris}}, \bibinfo {author} {\bibfnamefont
  {T.}~\bibnamefont {Loerting}}, \bibinfo {author} {\bibfnamefont {A.~Z.}\
  \bibnamefont {Panagiotopoulos}}, \bibinfo {author} {\bibfnamefont
  {J.}~\bibnamefont {Russo}}, \bibinfo {author} {\bibfnamefont {J.~A.}\
  \bibnamefont {Sellberg}}, \bibinfo {author} {\bibfnamefont {H.~E.}\
  \bibnamefont {Stanley}}, \bibinfo {author} {\bibfnamefont {H.}~\bibnamefont
  {Tanaka}}, \bibinfo {author} {\bibfnamefont {C.}~\bibnamefont {Vega}},
  \bibinfo {author} {\bibfnamefont {L.}~\bibnamefont {Xu}}, \ and\ \bibinfo
  {author} {\bibfnamefont {L.~G.~M.}\ \bibnamefont {Pettersson}},\ }\bibfield
  {title} {\enquote {\bibinfo {title} {Effect of hydrogen bonds on the
  thermodynamic behavior of liquid water},}\ }\href {\doibase
  10.1103/PhysRevLett.73.1632} {\bibfield  {journal} {\bibinfo  {journal}
  {Phys. Rev. Lett}\ }\textbf {\bibinfo {volume} {73}},\ \bibinfo {pages}
  {1632--1635} (\bibinfo {year} {1994})}\BibitemShut {NoStop}%
\bibitem [{\citenamefont {Duška}(2020)}]{Duska_Water_2020}%
  \BibitemOpen
  \bibfield  {author} {\bibinfo {author} {\bibfnamefont {M.}~\bibnamefont
  {Duška}},\ }\bibfield  {title} {\enquote {\bibinfo {title} {Water above the
  spinodal},}\ }\href {\doibase https://doi.org/10.1063/5.0006431} {\bibfield
  {journal} {\bibinfo  {journal} {J. Chem. Phys.}\ }\textbf {\bibinfo {volume}
  {152}},\ \bibinfo {pages} {174501} (\bibinfo {year} {2020})}\BibitemShut
  {NoStop}%
\bibitem [{\citenamefont {Caupin}\ and\ \citenamefont
  {Anisimov}(2019)}]{Caupin_Thermodynamics_2019}%
  \BibitemOpen
  \bibfield  {author} {\bibinfo {author} {\bibfnamefont {F.}~\bibnamefont
  {Caupin}}\ and\ \bibinfo {author} {\bibfnamefont {M.~A.}\ \bibnamefont
  {Anisimov}},\ }\bibfield  {title} {\enquote {\bibinfo {title} {Thermodynamics
  of supercooled and stretched water: Unifying two-structure description and
  liquid-vapor spinodal},}\ }\href {\doibase https://doi.org/10.1063/1.5100228}
  {\bibfield  {journal} {\bibinfo  {journal} {J. Chem. Phys.}\ }\textbf
  {\bibinfo {volume} {151}},\ \bibinfo {pages} {034503} (\bibinfo {year}
  {2019})}\BibitemShut {NoStop}%
\bibitem [{\citenamefont {Poole}\ \emph {et~al.}(1992)\citenamefont {Poole},
  \citenamefont {Sciortino}, \citenamefont {Essmann},\ and\ \citenamefont
  {Stanley}}]{Poole1992}%
  \BibitemOpen
  \bibfield  {author} {\bibinfo {author} {\bibfnamefont {P.~H.}\ \bibnamefont
  {Poole}}, \bibinfo {author} {\bibfnamefont {F.}~\bibnamefont {Sciortino}},
  \bibinfo {author} {\bibfnamefont {U.}~\bibnamefont {Essmann}}, \ and\
  \bibinfo {author} {\bibfnamefont {H.~E.}\ \bibnamefont {Stanley}},\
  }\bibfield  {title} {\enquote {\bibinfo {title} {Phase behavior of metastable
  water},}\ }\href {\doibase https://doi.org/10.1038/360324a0} {\bibfield
  {journal} {\bibinfo  {journal} {Nature}\ }\textbf {\bibinfo {volume} {360}},\
  \bibinfo {pages} {324--328} (\bibinfo {year} {1992})}\BibitemShut {NoStop}%
\bibitem [{\citenamefont {Holten}\ \emph {et~al.}(2014)\citenamefont {Holten},
  \citenamefont {Palmer}, \citenamefont {Poole}, \citenamefont {Debenedetti},\
  and\ \citenamefont {Anisimov}}]{Holten2001}%
  \BibitemOpen
  \bibfield  {author} {\bibinfo {author} {\bibfnamefont {V.}~\bibnamefont
  {Holten}}, \bibinfo {author} {\bibfnamefont {J.~C.}\ \bibnamefont {Palmer}},
  \bibinfo {author} {\bibfnamefont {P.~H.}\ \bibnamefont {Poole}}, \bibinfo
  {author} {\bibfnamefont {P.~G.}\ \bibnamefont {Debenedetti}}, \ and\ \bibinfo
  {author} {\bibfnamefont {M.~A.}\ \bibnamefont {Anisimov}},\ }\bibfield
  {title} {\enquote {\bibinfo {title} {Two-state thermodynamics of the st2
  model for supercooled water},}\ }\href {\doibase
  https://doi.org/10.1063/1.4867287} {\bibfield  {journal} {\bibinfo  {journal}
  {J. Chem. Phys.}\ }\textbf {\bibinfo {volume} {104502}} (\bibinfo {year}
  {2014}),\ https://doi.org/10.1063/1.4867287}\BibitemShut {NoStop}%
\bibitem [{\citenamefont {Debenedetti}, \citenamefont {Sciortino},\ and\
  \citenamefont {Zerze}(2020)}]{Debenedetti2020}%
  \BibitemOpen
  \bibfield  {author} {\bibinfo {author} {\bibfnamefont {P.~G.}\ \bibnamefont
  {Debenedetti}}, \bibinfo {author} {\bibfnamefont {F.}~\bibnamefont
  {Sciortino}}, \ and\ \bibinfo {author} {\bibfnamefont {G.~H.}\ \bibnamefont
  {Zerze}},\ }\bibfield  {title} {\enquote {\bibinfo {title} {Second critical
  point in two realistic models of water},}\ }\href {\doibase
  10.1126/science.abb9796} {\bibfield  {journal} {\bibinfo  {journal}
  {Science}\ }\textbf {\bibinfo {volume} {369}},\ \bibinfo {pages} {289--292}
  (\bibinfo {year} {2020})}\BibitemShut {NoStop}%
\bibitem [{\citenamefont {Biddle}\ \emph {et~al.}(2017)\citenamefont {Biddle},
  \citenamefont {Singh}, \citenamefont {Sparano}, \citenamefont {Ricci},
  \citenamefont {González}, \citenamefont {Valeriani}, \citenamefont
  {Abascal}, \citenamefont {Debenedetti}, \citenamefont {Anisimov}, ,\ and\
  \citenamefont {Caupin}}]{Biddle_Two_2017}%
  \BibitemOpen
  \bibfield  {author} {\bibinfo {author} {\bibfnamefont {J.~W.}\ \bibnamefont
  {Biddle}}, \bibinfo {author} {\bibfnamefont {R.~S.}\ \bibnamefont {Singh}},
  \bibinfo {author} {\bibfnamefont {E.~M.}\ \bibnamefont {Sparano}}, \bibinfo
  {author} {\bibfnamefont {F.}~\bibnamefont {Ricci}}, \bibinfo {author}
  {\bibfnamefont {M.~A.}\ \bibnamefont {González}}, \bibinfo {author}
  {\bibfnamefont {C.}~\bibnamefont {Valeriani}}, \bibinfo {author}
  {\bibfnamefont {J.~L.~F.}\ \bibnamefont {Abascal}}, \bibinfo {author}
  {\bibfnamefont {P.~G.}\ \bibnamefont {Debenedetti}}, \bibinfo {author}
  {\bibfnamefont {M.~A.}\ \bibnamefont {Anisimov}}, , \ and\ \bibinfo {author}
  {\bibfnamefont {F.}~\bibnamefont {Caupin}},\ }\bibfield  {title} {\enquote
  {\bibinfo {title} {Two-structure thermodynamics for the tip4p/2005 model of
  water covering supercooled and deeply stretched regions},}\ }\href
  {https://doi.org/10.1063/1.4973546} {\bibfield  {journal} {\bibinfo
  {journal} {J. Chem. Phys.}\ }\textbf {\bibinfo {volume} {146}},\ \bibinfo
  {pages} {034502} (\bibinfo {year} {2017})}\BibitemShut {NoStop}%
\bibitem [{\citenamefont {Debenedetti}(1998)}]{Debenedetti_One_1998}%
  \BibitemOpen
  \bibfield  {author} {\bibinfo {author} {\bibfnamefont {P.~G.}\ \bibnamefont
  {Debenedetti}},\ }\bibfield  {title} {\enquote {\bibinfo {title} {One
  substance, two liquids?}}\ }\href {\doibase https://doi.org/10.1038/32286}
  {\bibfield  {journal} {\bibinfo  {journal} {Nature}\ }\textbf {\bibinfo
  {volume} {392}},\ \bibinfo {pages} {127--128} (\bibinfo {year}
  {1998})}\BibitemShut {NoStop}%
\bibitem [{\citenamefont {Longo}\ and\ \citenamefont
  {Anisimov}(2022)}]{Longo2021}%
  \BibitemOpen
  \bibfield  {author} {\bibinfo {author} {\bibfnamefont {T.~J.}\ \bibnamefont
  {Longo}}\ and\ \bibinfo {author} {\bibfnamefont {M.~A.}\ \bibnamefont
  {Anisimov}},\ }\bibfield  {title} {\enquote {\bibinfo {title} {Phase
  transitions affected by natural and forceful molecular interconversion},}\
  }\href {\doibase https://doi.org/10.1063/5.0081180} {\bibfield  {journal}
  {\bibinfo  {journal} {J. Chem. Phys.}\ }\textbf {\bibinfo {volume} {156}},\
  \bibinfo {pages} {084502} (\bibinfo {year} {2022})}\BibitemShut {NoStop}%
\bibitem [{\citenamefont {Caupin}\ and\ \citenamefont
  {Anisimov}(2021)}]{Caupin2021}%
  \BibitemOpen
  \bibfield  {author} {\bibinfo {author} {\bibfnamefont {F.}~\bibnamefont
  {Caupin}}\ and\ \bibinfo {author} {\bibfnamefont {M.~A.}\ \bibnamefont
  {Anisimov}},\ }\bibfield  {title} {\enquote {\bibinfo {title} {Minimal
  microscopic model for liquid polyamorphism and waterlike anomalies},}\ }\href
  {\doibase 10.1103/PhysRevLett.127.185701} {\bibfield  {journal} {\bibinfo
  {journal} {Phys. Rev. Lett.}\ }\textbf {\bibinfo {volume} {127}},\ \bibinfo
  {pages} {185701} (\bibinfo {year} {2021})}\BibitemShut {NoStop}%
\bibitem [{\citenamefont {Sauer}\ and\ \citenamefont
  {Borst}(1967)}]{Sauer_Lambda_1967}%
  \BibitemOpen
  \bibfield  {author} {\bibinfo {author} {\bibfnamefont {G.~E.}\ \bibnamefont
  {Sauer}}\ and\ \bibinfo {author} {\bibfnamefont {L.~B.}\ \bibnamefont
  {Borst}},\ }\bibfield  {title} {\enquote {\bibinfo {title} {Lambda transition
  in liquid sulfur},}\ }\href {\doibase 10.1126/science.158.3808.1567}
  {\bibfield  {journal} {\bibinfo  {journal} {Science}\ }\textbf {\bibinfo
  {volume} {158}},\ \bibinfo {pages} {1567--1569} (\bibinfo {year}
  {1967})}\BibitemShut {NoStop}%
\bibitem [{\citenamefont {Bellissent}, \citenamefont {Descotes},\ and\
  \citenamefont {Pfeuty}(1994)}]{Bellissent_Sulfur_1994}%
  \BibitemOpen
  \bibfield  {author} {\bibinfo {author} {\bibfnamefont {R.}~\bibnamefont
  {Bellissent}}, \bibinfo {author} {\bibfnamefont {L.}~\bibnamefont
  {Descotes}}, \ and\ \bibinfo {author} {\bibfnamefont {P.}~\bibnamefont
  {Pfeuty}},\ }\bibfield  {title} {\enquote {\bibinfo {title} {Polymerization
  in liquid sulphur},}\ }\href {\doibase 10.1088/0953-8984/6/23a/031}
  {\bibfield  {journal} {\bibinfo  {journal} {J. Phys.: Condens. Matte}\
  }\textbf {\bibinfo {volume} {6}},\ \bibinfo {pages} {A211--A216} (\bibinfo
  {year} {1994})}\BibitemShut {NoStop}%
\bibitem [{\citenamefont {Kozhevnikov}\ \emph {et~al.}(2004)\citenamefont
  {Kozhevnikov}, \citenamefont {Payne}, \citenamefont {Olson}, \citenamefont
  {McDonald},\ and\ \citenamefont {Inglefield}}]{Kozhevnikov_Sulfur_2004}%
  \BibitemOpen
  \bibfield  {author} {\bibinfo {author} {\bibfnamefont {V.~F.}\ \bibnamefont
  {Kozhevnikov}}, \bibinfo {author} {\bibfnamefont {W.~B.}\ \bibnamefont
  {Payne}}, \bibinfo {author} {\bibfnamefont {J.~K.}\ \bibnamefont {Olson}},
  \bibinfo {author} {\bibfnamefont {C.~L.}\ \bibnamefont {McDonald}}, \ and\
  \bibinfo {author} {\bibfnamefont {C.~E.}\ \bibnamefont {Inglefield}},\
  }\bibfield  {title} {\enquote {\bibinfo {title} {Physical properties of
  sulfur near the polymerization transition},}\ }\href
  {https://doi.org/10.1063/1.1794031} {\bibfield  {journal} {\bibinfo
  {journal} {J. Chem. Phys.}\ }\textbf {\bibinfo {volume} {121}} (\bibinfo
  {year} {2004})}\BibitemShut {NoStop}%
\bibitem [{\citenamefont {Tobolsky}\ and\ \citenamefont
  {Eisenberg}(1959)}]{Tobolsky_Sulfur_1959}%
  \BibitemOpen
  \bibfield  {author} {\bibinfo {author} {\bibfnamefont {A.~V.}\ \bibnamefont
  {Tobolsky}}\ and\ \bibinfo {author} {\bibfnamefont {A.}~\bibnamefont
  {Eisenberg}},\ }\bibfield  {title} {\enquote {\bibinfo {title} {Equilibrium
  polymerization of sulfur},}\ }\href {\doibase
  https://doi.org/10.1021/ja01513a004} {\bibfield  {journal} {\bibinfo
  {journal} {J. Am. Chem. Soc.}\ }\textbf {\bibinfo {volume} {81}},\ \bibinfo
  {pages} {780--782} (\bibinfo {year} {1959})}\BibitemShut {NoStop}%
\bibitem [{\citenamefont {Eisenberg}\ and\ \citenamefont
  {Tobolsky}(1960)}]{Tobolsky_Selenium_1960}%
  \BibitemOpen
  \bibfield  {author} {\bibinfo {author} {\bibfnamefont {A.}~\bibnamefont
  {Eisenberg}}\ and\ \bibinfo {author} {\bibfnamefont {A.~V.}\ \bibnamefont
  {Tobolsky}},\ }\bibfield  {title} {\enquote {\bibinfo {title} {Equilibrium
  polymerization of selenium},}\ }\href {\doibase
  https://doi.org/10.1002/pol.1960.1204614703} {\bibfield  {journal} {\bibinfo
  {journal} {J. Pol. Sci.}\ }\textbf {\bibinfo {volume} {46}} (\bibinfo {year}
  {1960}),\ https://doi.org/10.1002/pol.1960.1204614703}\BibitemShut {NoStop}%
\bibitem [{\citenamefont {Shumovskyi}\ \emph {et~al.}(2022)\citenamefont
  {Shumovskyi}, \citenamefont {Longo}, \citenamefont {Buldyrev},\ and\
  \citenamefont {Anisimov}}]{Shumovskyi2022}%
  \BibitemOpen
  \bibfield  {author} {\bibinfo {author} {\bibfnamefont {N.~A.}\ \bibnamefont
  {Shumovskyi}}, \bibinfo {author} {\bibfnamefont {T.~J.}\ \bibnamefont
  {Longo}}, \bibinfo {author} {\bibfnamefont {S.~V.}\ \bibnamefont {Buldyrev}},
  \ and\ \bibinfo {author} {\bibfnamefont {M.~A.}\ \bibnamefont {Anisimov}},\
  }\bibfield  {title} {\enquote {\bibinfo {title} {Modeling fluid polyamorphism
  through a maximum-valence approach},}\ }\href {\doibase
  https://doi.org/10.1103/PhysRevE.106.015305} {\bibfield  {journal} {\bibinfo
  {journal} {Phys. Rev. E}\ }\textbf {\bibinfo {volume} {106}},\ \bibinfo
  {pages} {015305} (\bibinfo {year} {2022})}\BibitemShut {NoStop}%
\bibitem [{\citenamefont {Zaccarelli}\ \emph {et~al.}(2005)\citenamefont
  {Zaccarelli}, \citenamefont {Buldyrev}, \citenamefont {Nave}, \citenamefont
  {Moreno}, \citenamefont {Saika-Voivod}, \citenamefont {Sciortino},\ and\
  \citenamefont {Tartagliae}}]{Zaccarelli2005}%
  \BibitemOpen
  \bibfield  {author} {\bibinfo {author} {\bibfnamefont {E.}~\bibnamefont
  {Zaccarelli}}, \bibinfo {author} {\bibfnamefont {S.~V.}\ \bibnamefont
  {Buldyrev}}, \bibinfo {author} {\bibfnamefont {E.~L.}\ \bibnamefont {Nave}},
  \bibinfo {author} {\bibfnamefont {A.~J.}\ \bibnamefont {Moreno}}, \bibinfo
  {author} {\bibfnamefont {I.}~\bibnamefont {Saika-Voivod}}, \bibinfo {author}
  {\bibfnamefont {F.}~\bibnamefont {Sciortino}}, \ and\ \bibinfo {author}
  {\bibfnamefont {P.}~\bibnamefont {Tartagliae}},\ }\bibfield  {title}
  {\enquote {\bibinfo {title} {Model for reversible colloidal gelation},}\
  }\href {\doibase 10.1103/PhysRevLett.94.218301} {\bibfield  {journal}
  {\bibinfo  {journal} {Phys. Rev. Lett.}\ }\textbf {\bibinfo {volume} {94}},\
  \bibinfo {pages} {218301} (\bibinfo {year} {2005})}\BibitemShut {NoStop}%
\bibitem [{\citenamefont {Speedy}\ and\ \citenamefont
  {Debenedetti}(1994)}]{Debenedetti}%
  \BibitemOpen
  \bibfield  {author} {\bibinfo {author} {\bibfnamefont {R.~J.}\ \bibnamefont
  {Speedy}}\ and\ \bibinfo {author} {\bibfnamefont {P.~G.}\ \bibnamefont
  {Debenedetti}},\ }\bibfield  {title} {\enquote {\bibinfo {title} {The entropy
  of a network crystal, fluid and glass},}\ }\href {\doibase
  10.1080/00268979400100161} {\bibfield  {journal} {\bibinfo  {journal} {Mol.
  Phys.}\ }\textbf {\bibinfo {volume} {81}},\ \bibinfo {pages} {237--249}
  (\bibinfo {year} {1994})}\BibitemShut {NoStop}%
\bibitem [{\citenamefont {Speedy}\ and\ \citenamefont
  {Debenedetti}(1996)}]{Debenedetti2}%
  \BibitemOpen
  \bibfield  {author} {\bibinfo {author} {\bibfnamefont {R.~J.}\ \bibnamefont
  {Speedy}}\ and\ \bibinfo {author} {\bibfnamefont {P.~G.}\ \bibnamefont
  {Debenedetti}},\ }\bibfield  {title} {\enquote {\bibinfo {title} {The
  distribution of tetravalent network glasses},}\ }\href {\doibase
  10.1080/00268979609484512} {\bibfield  {journal} {\bibinfo  {journal} {Mol.
  Phys.}\ }\textbf {\bibinfo {volume} {88}},\ \bibinfo {pages} {1293--1316}
  (\bibinfo {year} {1996})}\BibitemShut {NoStop}%
\bibitem [{\citenamefont {Wigner}\ and\ \citenamefont
  {Huntington}(1935)}]{Wigner1935}%
  \BibitemOpen
  \bibfield  {author} {\bibinfo {author} {\bibfnamefont {E.}~\bibnamefont
  {Wigner}}\ and\ \bibinfo {author} {\bibfnamefont {H.~B.}\ \bibnamefont
  {Huntington}},\ }\bibfield  {title} {\enquote {\bibinfo {title} {On the
  possibility of a metallic modification of hydrogen},}\ }\href {\doibase
  https://doi.org/10.1063/1.1749590} {\bibfield  {journal} {\bibinfo  {journal}
  {The Journal of Chemical Physics}\ }\textbf {\bibinfo {volume} {3}} (\bibinfo
  {year} {1935}),\ https://doi.org/10.1063/1.1749590}\BibitemShut {NoStop}%
\bibitem [{\citenamefont {Giguere}(1984)}]{Giguere1984}%
  \BibitemOpen
  \bibfield  {author} {\bibinfo {author} {\bibfnamefont {P.~A.}\ \bibnamefont
  {Giguere}},\ }\bibfield  {title} {\enquote {\bibinfo {title} {Bifurcated
  hydrogen bonds in water},}\ }\href {\doibase 10.1002/jrs.1250150513}
  {\bibfield  {journal} {\bibinfo  {journal} {The Journal of Raman
  Spectroscopy}\ }\textbf {\bibinfo {volume} {15}},\ \bibinfo {pages}
  {354--359} (\bibinfo {year} {1984})}\BibitemShut {NoStop}%
\bibitem [{\citenamefont {Alder}\ and\ \citenamefont
  {Wainwright}(1959)}]{Alder1959}%
  \BibitemOpen
  \bibfield  {author} {\bibinfo {author} {\bibfnamefont {B.~J.}\ \bibnamefont
  {Alder}}\ and\ \bibinfo {author} {\bibfnamefont {T.~E.}\ \bibnamefont
  {Wainwright}},\ }\bibfield  {title} {\enquote {\bibinfo {title} {Studies in
  molecular dynamics. i. general method},}\ }\href {\doibase
  https://doi.org/10.1063/1.1730376} {\bibfield  {journal} {\bibinfo  {journal}
  {J. Chem. Phys.}\ }\textbf {\bibinfo {volume} {31}},\ \bibinfo {pages} {459}
  (\bibinfo {year} {1959})}\BibitemShut {NoStop}%
\bibitem [{\citenamefont {Rapaport}(2004)}]{Rapaport2004}%
  \BibitemOpen
  \bibfield  {author} {\bibinfo {author} {\bibfnamefont {D.~C.}\ \bibnamefont
  {Rapaport}},\ }\href@noop {} {\emph {\bibinfo {title} {The Art of Molecular
  Dynamics Simulation}}},\ \bibinfo {edition} {2nd}\ ed.\ (\bibinfo
  {publisher} {Cambridge University Press},\ \bibinfo {address} {Cambridge,
  UK},\ \bibinfo {year} {2004})\BibitemShut {NoStop}%
\bibitem [{\citenamefont {Buldyrev}(2009)}]{Buldyrev_Application_2008}%
  \BibitemOpen
  \bibfield  {author} {\bibinfo {author} {\bibfnamefont {S.}~\bibnamefont
  {Buldyrev}},\ }\bibfield  {title} {\enquote {\bibinfo {title} {Application of
  discrete molecular dynamics to protein folding and aggregation},}\ }in\ \href
  {\doibase https://doi.org/10.1007/978-3-540-78765-5_5} {\emph {\bibinfo
  {booktitle} {Aspects of Physical Biology}}},\ \bibinfo {series} {Lecture
  Notes in Physics}, Vol.\ \bibinfo {volume} {752},\ \bibinfo {editor} {edited
  by\ \bibinfo {editor} {\bibfnamefont {G.}~\bibnamefont {Franzese}}\ and\
  \bibinfo {editor} {\bibfnamefont {M.}~\bibnamefont {Rubi}}}\ (\bibinfo
  {publisher} {Springer-Verlag},\ \bibinfo {address} {Berlin, Heidelberg},\
  \bibinfo {year} {2009})\ pp.\ \bibinfo {pages} {97--132}\BibitemShut
  {NoStop}%
\bibitem [{\citenamefont {Berendsen}\ \emph {et~al.}(1984)\citenamefont
  {Berendsen}, \citenamefont {Postma}, \citenamefont {van Gunsteren},
  \citenamefont {DiNola},\ and\ \citenamefont {Haak}}]{Berendsen1984}%
  \BibitemOpen
  \bibfield  {author} {\bibinfo {author} {\bibfnamefont {H.~J.~C.}\
  \bibnamefont {Berendsen}}, \bibinfo {author} {\bibfnamefont {J.~P.~M.}\
  \bibnamefont {Postma}}, \bibinfo {author} {\bibfnamefont {W.~F.}\
  \bibnamefont {van Gunsteren}}, \bibinfo {author} {\bibfnamefont
  {A.}~\bibnamefont {DiNola}}, \ and\ \bibinfo {author} {\bibfnamefont {J.~R.}\
  \bibnamefont {Haak}},\ }\bibfield  {title} {\enquote {\bibinfo {title}
  {Molecular-dynamics with coupling to an external bath},}\ }\href@noop {}
  {\bibfield  {journal} {\bibinfo  {journal} {J. Chem. Phys.}\ }\textbf
  {\bibinfo {volume} {81}},\ \bibinfo {pages} {3684--3690} (\bibinfo {year}
  {1984})}\BibitemShut {NoStop}%
\bibitem [{\citenamefont {Skibinsky}\ \emph {et~al.}(2004)\citenamefont
  {Skibinsky}, \citenamefont {Buldyrev}, \citenamefont {Franzese},
  \citenamefont {Malescio},\ and\ \citenamefont {Stanley}}]{Skibinsky2004}%
  \BibitemOpen
  \bibfield  {author} {\bibinfo {author} {\bibfnamefont {A.}~\bibnamefont
  {Skibinsky}}, \bibinfo {author} {\bibfnamefont {S.~V.}\ \bibnamefont
  {Buldyrev}}, \bibinfo {author} {\bibfnamefont {G.}~\bibnamefont {Franzese}},
  \bibinfo {author} {\bibfnamefont {G.}~\bibnamefont {Malescio}}, \ and\
  \bibinfo {author} {\bibfnamefont {H.~E.}\ \bibnamefont {Stanley}},\
  }\bibfield  {title} {\enquote {\bibinfo {title} {Liquid-liquid phase
  transitions for soft-core attractive potentials},}\ }\href {\doibase
  10.1103/PhysRevE.69.061206} {\bibfield  {journal} {\bibinfo  {journal} {Phys.
  Rev. E}\ }\textbf {\bibinfo {volume} {69}},\ \bibinfo {pages} {61206--15}
  (\bibinfo {year} {2004})}\BibitemShut {NoStop}%
\bibitem [{\citenamefont {Fried}, \citenamefont {Longo},\ and\ \citenamefont
  {An}(2022)}]{Fried2022}%
  \BibitemOpen
  \bibfield  {author} {\bibinfo {author} {\bibfnamefont {N.~R.}\ \bibnamefont
  {Fried}}, \bibinfo {author} {\bibfnamefont {T.~J.}\ \bibnamefont {Longo}}, \
  and\ \bibinfo {author} {\bibfnamefont {M.~A.}\ \bibnamefont {An}},\
  }\bibfield  {title} {\enquote {\bibinfo {title} {Modeling fluid polyamorphism
  through a maximum-valence approach},}\ }\href {\doibase https://doi:
  10.1063/5.0107043} {\bibfield  {journal} {\bibinfo  {journal} {J. Chem.
  Phys.}\ }\textbf {\bibinfo {volume} {157}},\ \bibinfo {pages} {101101}
  (\bibinfo {year} {2022})}\BibitemShut {NoStop}%
\bibitem [{\citenamefont {Poole}\ \emph {et~al.}(2016)\citenamefont {Poole},
  \citenamefont {Sciortino}, \citenamefont {Grande}, \citenamefont {Stanley},\
  and\ \citenamefont {Angell}}]{Poole1994}%
  \BibitemOpen
  \bibfield  {author} {\bibinfo {author} {\bibnamefont {Poole}}, \bibinfo
  {author} {\bibnamefont {Sciortino}}, \bibinfo {author} {\bibnamefont
  {Grande}}, \bibinfo {author} {\bibnamefont {Stanley}}, \ and\ \bibinfo
  {author} {\bibnamefont {Angell}},\ }\bibfield  {title} {\enquote {\bibinfo
  {title} {Water: A tale of two liquids},}\ }\href {\doibase
  10.1021/acs.chemrev.5b00750} {\bibfield  {journal} {\bibinfo  {journal}
  {Chem. Rev.}\ }\textbf {\bibinfo {volume} {116}},\ \bibinfo {pages}
  {7463--7500} (\bibinfo {year} {2016})}\BibitemShut {NoStop}%
\bibitem [{\citenamefont {Kim}\ \emph {et~al.}(2020)\citenamefont {Kim},
  \citenamefont {Amann-Winkel}, \citenamefont {Giovambattista}, \citenamefont
  {Späh}, \citenamefont {Perakis}, \citenamefont {Pathak}, \citenamefont
  {Parada}, \citenamefont {Yang}, \citenamefont {Mariedahl}, \citenamefont
  {Eklund}, \citenamefont {Lane}, \citenamefont {You}, \citenamefont {Jeong},
  \citenamefont {Weston}, \citenamefont {Lee}, \citenamefont {Eom},
  \citenamefont {Kim}, \citenamefont {Park}, \citenamefont {Chun},
  \citenamefont {Poole},\ and\ \citenamefont {Nilsson}}]{Kim2020}%
  \BibitemOpen
  \bibfield  {author} {\bibinfo {author} {\bibfnamefont {K.~H.}\ \bibnamefont
  {Kim}}, \bibinfo {author} {\bibfnamefont {K.}~\bibnamefont {Amann-Winkel}},
  \bibinfo {author} {\bibfnamefont {N.}~\bibnamefont {Giovambattista}},
  \bibinfo {author} {\bibfnamefont {A.}~\bibnamefont {Späh}}, \bibinfo
  {author} {\bibfnamefont {F.}~\bibnamefont {Perakis}}, \bibinfo {author}
  {\bibfnamefont {H.}~\bibnamefont {Pathak}}, \bibinfo {author} {\bibfnamefont
  {M.~L.}\ \bibnamefont {Parada}}, \bibinfo {author} {\bibfnamefont
  {C.}~\bibnamefont {Yang}}, \bibinfo {author} {\bibfnamefont {D.}~\bibnamefont
  {Mariedahl}}, \bibinfo {author} {\bibfnamefont {T.}~\bibnamefont {Eklund}},
  \bibinfo {author} {\bibfnamefont {T.~J.}\ \bibnamefont {Lane}}, \bibinfo
  {author} {\bibfnamefont {S.}~\bibnamefont {You}}, \bibinfo {author}
  {\bibfnamefont {S.}~\bibnamefont {Jeong}}, \bibinfo {author} {\bibfnamefont
  {M.}~\bibnamefont {Weston}}, \bibinfo {author} {\bibfnamefont {J.~H.}\
  \bibnamefont {Lee}}, \bibinfo {author} {\bibfnamefont {I.}~\bibnamefont
  {Eom}}, \bibinfo {author} {\bibfnamefont {M.}~\bibnamefont {Kim}}, \bibinfo
  {author} {\bibfnamefont {J.}~\bibnamefont {Park}}, \bibinfo {author}
  {\bibfnamefont {S.~H.}\ \bibnamefont {Chun}}, \bibinfo {author}
  {\bibfnamefont {P.~H.}\ \bibnamefont {Poole}}, \ and\ \bibinfo {author}
  {\bibfnamefont {A.}~\bibnamefont {Nilsson}},\ }\bibfield  {title} {\enquote
  {\bibinfo {title} {Experimental observation of the liquid-liquid transition
  in bulk supercooled water under pressure},}\ }\href {\doibase
  10.1126/science.abb9385} {\bibfield  {journal} {\bibinfo  {journal}
  {Science}\ }\textbf {\bibinfo {volume} {370}},\ \bibinfo {pages} {978--982}
  (\bibinfo {year} {2020})}\BibitemShut {NoStop}%
\bibitem [{\citenamefont {Bolhuis}, \citenamefont {Hagen},\ and\ \citenamefont
  {Frenkel}(1984)}]{Bolhius1984}%
  \BibitemOpen
  \bibfield  {author} {\bibinfo {author} {\bibfnamefont {P.}~\bibnamefont
  {Bolhuis}}, \bibinfo {author} {\bibfnamefont {M.}~\bibnamefont {Hagen}}, \
  and\ \bibinfo {author} {\bibfnamefont {D.}~\bibnamefont {Frenkel}},\
  }\bibfield  {title} {\enquote {\bibinfo {title} {Isostructural solid-solid
  transition in crystalline systems with short-ranged interaction},}\ }\href
  {\doibase 10.1103/PhysRevE.50.4880} {\bibfield  {journal} {\bibinfo
  {journal} {Phys. Rev. E}\ }\textbf {\bibinfo {volume} {50}} (\bibinfo {year}
  {1984}),\ 10.1103/PhysRevE.50.4880}\BibitemShut {NoStop}%
\end{thebibliography}%

\end{document}